\documentclass[fleqn,usenatbib]{mnras}

\usepackage{newtxtext,newtxmath}
\usepackage{xcolor}

\usepackage[T1]{fontenc}
\usepackage{ae,aecompl}

\usepackage{graphicx}	
\usepackage{amsmath}	
\usepackage{amssymb}	

\usepackage{soul}	
\usepackage{xcolor}

\title[Diffuse emission from starburst nuclei]{Contribution of starburst nuclei to the diffuse gamma-ray and neutrino flux}

\author[E. Peretti et al.]{Enrico Peretti$^{1,2}$ \thanks{E-mail: enrico.peretti@gssi.it},
Pasquale Blasi$^{1,2}$ 
,
Felix Aharonian$^{3,4,1}$ 
,  \newauthor Giovanni Morlino$^{5}$ and 
Pierre Cristofari$^{1,2}$ 
\\
\\
$^{1}$Gran Sasso Science Institute, Viale F. Crispi 7, 67100 L'Aquila, Italy \\
$^{2}$INFN/Laboratori Nazionali del Gran Sasso, via G. Acitelli 22, 67100 Assergi (AQ), Italy \\
$^{3}$Dublin Institute for Advanced Studies, 31 Fitzwilliam Place, D04 C932 Dublin 2, Ireland \\
$^{4}$Max-Planck-Institute f{\"u}r Kernphysik, Saupfercheckweg 1, D-69117 Heidelberg, Germany \\
$^{5}$INAF/Osservatorio Astrofisico di Arcetri, L.go E. Fermi 5, 50125 Firenze, Italy \\
}

\date{Accepted XXX. Received YYY; in original form ZZZ}

\pubyear{2019}

\begin{document}
\label{firstpage}
\pagerange{\pageref{firstpage}--\pageref{lastpage}}
\maketitle

\begin{abstract}
In nuclei of starburst galaxies, the combination of an enhanced rate of supernova explosions and a high gas density suggests that cosmic rays can be efficiently produced, and that most of them lose their energy before escaping these regions, resulting in a large flux of secondary products, including neutrinos. Although the flux inferred from an individual starburst region is expected to be well below the sensitivity of current neutrino telescopes, such sources may provide a substantial contribution to the diffuse neutrino flux measured by IceCube.
Here we compute the gamma--ray and neutrino flux due to starburst galaxies based on a physical model of cosmic ray transport in a starburst nucleus, and accounting for the redshift evolution of the number density of starburst sources as inferred from recent measurements of the star formation rate. The model accounts for gamma--ray absorption both inside the sources and in the intergalactic medium. The latter process is responsible for electromagnetic cascades, which also contribute to the diffuse gamma--ray background at lower energies. The conditions for acceleration of cosmic ray protons up to energies exceeding $ \sim 10 \, \rm PeV$ in starburst regions, necessary for the production of PeV neutrinos, are investigated in a critical way. We show that starburst nuclei can account for the diffuse neutrino flux above $\sim 200 \, \rm TeV$, thereby producing $\lesssim 40 \%$ of the extragalactic diffuse gamma--ray background. Below $\sim 200 \, \rm TeV$, the flux from starburst appears to be somewhat lower than the observed one, where both the Galactic contribution and the flux of atmospheric neutrinos may account for the difference. 
\end{abstract}




\section{Introduction}

Starburst galaxies (SBGs) are powerful cosmic ray (CR) factories characterised by intense star formation rate (SFR) and extreme properties of their interstellar medium (ISM) \citep[see i.e.][]{Gao_Solomon_2004,Mannucci_etal_2003,ForsterSchreiber_2001,Thompson_etal_2006}. The enhanced star forming activity leads to an increased rate of supernova (SN) explosions and most likely a high rate of  CR production and highly turbulent medium, which in turn may effectively confine CRs for times exceeding the loss time. 

Starburst episodes often occur in relatively small regions, called starburst nuclei (SBNi), located in the central galactic regions \citep[see also][]{Loose-burst-SF-SBN}. Several observations suggest that these regions host a prominent population of non--thermal particles which emit radiation, both of leptonic origin, typically extending from radio to hard X--rays \citep[for the case of NGC253 see i.e.][]{Radio_Williams_Bower_2010,Carilli_2006,Wik_NGC253_2014}, and of hadronic origin, at high and very--high energies (VHE) \citep[see i.e.][]{Ackermann_Fermi_2012,Peng:2016nsx,Abdalla:2018nlz}. At VHE, $\gamma \gamma$ absorption due to the presence of an intense far infrared to optical (FIR--OPT) thermal background radiation, is expected to reduce the gamma radiation leaving the compact nuclei.

The recent discovery by the IceCube collaboration of a diffuse astrophysical neutrino flux \citep{First_Ice_nu}, of probable extragalactic origin, has renewed the interest in SBNi as CR and neutrino factories, one of the reasons being the possibility of producing relatively hard neutrino spectra: both in the case that CR transport in SBNi is dominated by energy losses and advection, the equilibrium spectrum of CRs is expected to have a spectral shape similar to that injected at the sources.

Many authors have recently modelled the CR spectrum in SBGs on the basis of their associated multiwavelength spectra \citep[][]{Paglione_1996, Torres_2004, Persic_2008, Rephaeli_2011, Lacki_Thompson_2013, Yoast-Hull_M82_2013, Wang_2018,Peretti:2018tmo}. From this bulk of work, consensus emerged on the fact that CR electrons typically lose their energy effectively inside the SBN, making the assumption of calorimetry well justified. Protons also behave approximately in a calorimetric way, to an extent that depends on the diffusive properties of the ISM and on the speed of the winds that the SBN ejects. 

In general, gamma rays set an upper limit for the associated neutrino flux, but this condition can be partially relaxed if gamma--gamma absorption inside the source is efficient. 
The hard injection spectrum of protons and the high target density, together with the efficient absorption of VHE gamma rays, make SBGs promising diffuse neutrino emitters \citep[see for related discussions][]{RomeroTorres_2003, DeCeaDelPozo_2009, Yoast_Hull_Thesis} with a possible major contribution from ultraluminous infrared galaxies (ULIRGs) \citep{HE_2013}. 
On the other hand, in spite of their unique properties, their large distances ($\gtrsim \, \rm Mpc $) make the detection by current instruments challenging. In fact, the number of SBGs detected in gamma rays is less than a dozen \citep[see][]{Fermi-4FGL}, and with the exception of Arp220 ($\sim 77$ Mpc), they are observed only in the vicinity of our Galaxy ($< 20$ Mpc).
However, the larger starburst activity at high redshift \citep[][]{Madau:2014bja} make SBNi potentially good candidate sources of the diffuse high energy neutrino background \citep{Loeb-Waxman:2006, Thompson-Waxman_2006, Stecker:2006vz}. 

The production of neutrinos in hadronic collisions is inevitably accompanied by the production of high energy (HE) photons, hence SBGs should also contribute to the extragalactic diffuse gamma--ray background (EGB) and one should, of course, check that the requirements necessary to fit the neutrino flux do not lead to overproduction of the observed EGB. 

Even before the discovery of the astrophysical neutrino flux, it was pointed out by \citet{Lacki_etal_2011} that the diffuse gamma--ray background due to SBGs is comparable, within a factor of a few, to the EGB as measured by Fermi--LAT \citep{FirstFermiDiff}. Following this line of thought, \citet{Murase:2013rfa} showed that neutrino spectrum from SBGs should be harder than 2.1--2.2 in order to not exceed the EGB. The constraint imposed by the measured EGB became more stringent after \citet{Fermi-LAT>50} showed that a fraction $86 \% ^{+16 \%}_{-14 \%} $ of the detected EGB should be due to unresolved blazars, leaving less room for the contribution of SBGs. A slightly smaller contribution from balzars, $68\% ^{+9 \%}_{-8 \%} $, was recently derived by \citet{Lisanti_2016} using a similar analysis. 
This constraint led some authors \citep{Bechtol-Ahlers:2015, Sudoh:2018} to claim that only part of the observed neutrino flux could be attributed to star forming galaxies. 

Other authors \citep{Liu_Aharonian_2014, Chang:2014hua, Chang:2014sua, Tamborra-Ando-Murase:2014} reached a different conclusion, reassessing that SBGs can be responsible for the whole IceCube neutrino flux. In particular, \citet{Palladino-starburst:2018} showed that the flux of neutrinos from SBGs is compatible with the  through-going muon neutrino flux \citep[][]{IceCube_muons}, which might represent the cleanest neutrino sample of extragalactic origin accounting for their northern hemisphere arrival direction \citep[see][for a detailed discussion]{Ahlers:2018fkn}. 
As we discuss below, this discrepancy may reflect the urge to explain the neutrino flux below $\sim 200 \, \rm TeV$, or rather allowing for other sources in such energy range.

The ambiguity of these statements is partly understandable since together with blazars, SBGs have long been considered as main contributors to the EGB \citep{Soltan-SFGs-Gamma}. In fact, some active galactic nuclei (AGN) show starburst features as well \citep{Levenson:2000qd,Imanishi:2003hi,Yoast-Hull_AGN?}. Hence, depending on details of the calculations, the contribution of SBGs to the EGB can be saturated or close to be such, or leave enough room for the diffuse neutrino background to be explained.


An important step forward toward clarifying the situation is to have a physical understanding of the conditions for CR acceleration and transport in SBNi and propagation of gamma radiation inside the extreme environment typical of starburst nuclei. 
Furthermore, the flux of neutrinos and the associated flux of cascade photons depends on the cosmological evolution of the starburst activity with redshift. The latter aspect is usually accounted for by linking the evolution of SBGs to the history of star formation of the Universe, using as a probe the infrared luminosity function measured at different redshifts \citep[ e.g.][]{Gruppioni_2013_Lumin_func, Hopkins:2006bw, Yuksel-GRB_SFR}. In such an approach there is, however, the intrinsic uncertainty connected to the choice of the smallest value of IR luminosity at which a SBG can be still considered to be an efficient gamma--ray and neutrino factory. 

Here we apply our previous modelling of CR transport in SBNi \citep[][hereafter P18]{Peretti:2018tmo} and propose an operative definition of starburst activity aimed at describing the gamma--ray production of these sources. 
We define the SBN as a region with intense star formation that is also able to confine CRs on time scales exceeding the loss time of hadronic CRs inside the same region. We do so by finding a connection between this condition and the IR luminosity of the SBN, namely its star formation rate (SFR). 

Such a definition allows us to put a physically motivated lower bound to the SBG population independent of their redshift and to correctly count the number of sources that contribute to the diffuse gamma--ray and neutrino flux using the Star Formation Rate Function (SFRF) approach. Such approach is helpful in disentangling the  contamination due to Active Galactic Nuclei (AGN)~ \citep{Gruppioni_2015_2}.

The production of neutrinos with energy $\gtrsim$ PeV requires CR acceleration up to energy $\gtrsim 100$ PeV. While this assumption is typically adopted throughout the existing literature, we think it is of crucial importance to assess how credible it is that CR sources inside SBNi or around them can in fact energize CRs up to such high energies. We discuss this point in the light of current knowledge of CR acceleration. 

We find that SBNi can provide a good description of the neutrino flux observed by IceCube above $\sim 200$ TeV without exceeding the constraints coming from the EGB, while accounting for virtually all the diffuse gamma--ray background of non-blazar origin above 50 GeV. The first IceCube data points,  below $\sim 200$ TeV, correspond to a flux that is about $\sim 2$ times larger than the flux predicted from SBNi, therefore suggesting that at least 50\% may originate from sources other than SBNi, such as: normal galaxies~\citep{Bechtol-Ahlers:2015}, our own Galaxy~\citep{NERONOV201660}, a confinement region around our Galaxy~\citep{Taylor_Neutrinos,Gal-neutrinos,Blasi_Amato19}, AGN~\citep{Neutrinos-opaque-sources,TXSa,TXSb}
or possibly hypernovae and hidden CR accelerators \citep{Hypernovae-murase,Murase-hidden-CR-ACC}.
In the energy region below 200 TeV, atmospheric neutrino contamination cannot yet be ruled out~\citep{Mascaretti_Blasi,Mascaretti_Vissani_prompt}.


The article is organized as follows: in Sec.~\ref{Sezione1} we briefly describe the model of particle acceleration in SBNi as detailed in P18 and the phenomena associated to the photon propagation; in Sec.~\ref{Sec_counting_prototype} we describe our assumptions on the number count of sources and on the prototype--based approach; in Sec.~\ref{Section_results} we report our results and discuss their physical implications in Sec.~\ref{discussion} with special attention to the maximum energy.
Our conclusions are summarized in Sec.~\ref{Section_conclusions}.
In our calculations we assume a standard $\Lambda$CDM cosmology with $\Omega_{\rm M}=0.31$, $\Omega_{\Lambda}=0.69$ and $H_0= 67.74 \rm \, km \, s^{-1} \, Mpc^{-1}$.

\section{Cosmic rays in SBNi and associated emission}
\label{Sezione1}

Following P18 we describe the transport of CRs in an SBN using a simple model in which diffusion and losses occur at the same rate at any location in the nucleus and advection is treated as a process that takes particles away from the production region. These assumptions unavoidably lead to a leaky-box treatment of transport where the CR distribution function, $f$, is given by the following equation:
\begin{equation}
\label{Transport}
    Q(p)= \frac{f(p)}{\tau_{\rm loss}(p)}+\frac{f(p)}{\tau_{\rm adv}(p)}+\frac{f(p)}{\tau_{\rm diff}(p)} \, .
\end{equation}
$Q$ is the injection rate per unit volume, assumed to be a power law in momentum, with index $\alpha$, times an exponential cut off $\text{exp}(-p/p_{p, \text{ max} })$ and $ \text{exp}(-p^2/p^2_{e, \text{ max}})$ respectively for protons and electrons. In this section, as in P18, we assume $p_{e, \text{ max}}= 10$ TeV c$^{-1}$, whereas we consider a range of possible values for $p_{p, \text{ max}}$  from $1 \, \rm PeV \, c^{-1}$ to a few $10^2 \, \rm PeV \, c^{-1}$, and we discuss implications below.

The injection due to SNe reads:
\begin{equation}
\label{SNe inj}
    Q(p)= \mathcal{R}_{\rm SN} \mathcal{N}(p)/V_{\rm SBN}
\end{equation}
and is normalized by imposing: 
\begin{equation}
    \int^{\infty}_{0} 4 \pi p^2 T(p) \mathcal{N}(p) \; dp = \xi_{\rm CR}E_{\rm SN} \, .
    \label{Inj_norm}
\end{equation}
Here $T(p)$ is the particle kinetic energy, $\xi_{\rm CR}$ is the acceleration efficiency, $E_{\rm SN}$ is the kinetic energy of the supernova ejecta. In the following we adopt $\xi_{\rm CR}= 10\%$ and $E_{\rm SN}=10^{51}$ erg as reference values. 
Primary electrons are injected assuming an electron-to-proton ratio $K_{ep}=1/50$, following what is inferred for our Galaxy. The solution of Eq. \eqref{Transport} can be written as:
\begin{equation}
    f(p)= Q(p) \tau(p) = \frac{Q(p)}{\tau_{\rm loss}^{-1}(p)+\tau_{\rm adv}^{-1}(p)+\tau_{\rm diff}^{-1}(p)}
    \label{Transport_solution}
\end{equation}
where $\tau(p)= \big[ \tau_{\rm loss}^{-1}(p)+\tau_{\rm adv}^{-1}(p)+\tau_{\rm diff}^{-1}(p) \big]^{-1}$ is the typical lifetime of a particle of momentum $p$ inside the source.

The quantities $\tau_{\rm loss}$, $\tau_{\rm adv}$ and $\tau_{\rm diff}$ are the typical timescales for energy losses, advection and diffusion, respectively. 
In the case of protons, the mechanisms responsible for energy losses are pp collisions, Coulomb interactions and ionization. Electrons lose energy through ionization, bremsstrahlung (BREM), synchrotron (SYN) and inverse Compton (IC) scattering. The escape of particles from the SBN is regulated by the advection due to the starburst wind ($\tau_{\rm adv}= R/v_{\rm wind}$) and by diffusion ($\tau_{\rm diff}(p)= R^2/D(p)$), described by the diffusion coefficient $D(p)$. 

As discussed in P18, the high rate of SN explosions in the SBN is likely to produce a high level of turbulence, which is expected to reflect in a small diffusion coefficient. A theory of diffusion in strong turbulence was developed by \cite{subedi}: the transport in these conditions can be approximated with a diffusion coefficient that has a functional shape 
\begin{equation}
\label{Difff1}
    D(p)= \frac{r_L(p) v(p)}{3 \mathcal{F}(k)},
\end{equation}
where $r_{\rm L}$ is the Larmor radius, $v$ the particle velocity and $\mathcal{F}(k)$ is the energy density of the turbulent magnetic field per unit of logarithmic wavenumber $k$, normalized to unity at the wavenumber $k_0$ corresponding to the turbulence injection scale. We assume a Kolmogorov--like spectrum, hence $\mathcal{F}(k) \propto k^{-d+1}$ with $d=5/3$. Moreover the typical injection length for turbulence $k_0^{-1}= L_0 = 1$~pc. 
Different assumptions on the diffusion coefficient have been already explored in P18 where we found that a wide range of diffusion coefficients leads to equally good CR confinement.
%

Although similar to the expression derived in quasi-linear theory, the physical justification for Eq. \ref{Difff1} is that at low energies particles move locally under the action of a magnetic field dominated by the largest spatial scales, provided the power spectrum is steep enough. 

The flux of neutrinos and gamma rays produced by CR protons through inelastic collisions is computed following the approach of \cite{Kelner_Aharonian_2006_proton-proton}, as summarized in P18.

An essential ingredient for our calculation of the gamma--ray emission from SBNi is the absorption of photons due to pair production on intense FIR--OPT thermal background of the SB environment. 
This effect is described in terms of an absorption coefficient $\eta_{\gamma \gamma}(E)= \int \sigma_{\gamma \gamma}(E,E') n_{\rm bkg}(E') dE'$, as used in the integration of the radiative transfer equation \citep[][]{Rybicki_Lightman} (see P18 for additional details). 
It is important to stress that inside the SBN pair production leads to effective suppression of the gamma--ray flux rather than an electromagnetic cascade. This is due to the fact that electron--positron pairs produced in the scattering lose energy rapidly through SYN emission in the intense magnetic field. This effect reduces considerably the gamma--ray flux at energies high enough to start an electromagnetic cascade during propagation on cosmological distances.

The flux of gamma rays and neutrinos at the Earth can be easily calculated from the fluxes produced at the sources. 
This calculation is detailed in Appendix~\ref{app:calculation}.
While traversing the IGM, HE gamma rays interact with low--energy photons of the Cosmic Microwave Background (CMB) and (direct or reprocessed) starlight, known as Extragalactic Background Light (EBL), leading to an electromagnetic cascade \citep[][]{Berezinsky_Smirnov}. A simple derivation of the spectrum of the cascade, which is, in good approximation, universal, is provided in Appendix~\ref{Appendix C}. 
This analytical description of the cascade \citep[][]{Berezinsky_em_cascade:2016} can be applied to cases in which the cascade is fully developed and the spectrum of the background photons can be approximated as a $\delta$--function in energy. In our case we model the background photon field with two $\delta$--functions, one for the CMB, at typical energy $\epsilon_{\rm CMB}$, and one for the EBL, at typical energy $\epsilon_{\rm EBL}$. 
In general, both contributions depend on redshift but, while the CMB dependence is well known, for the EBL the situation is somewhat more model dependent, even if a mild dependence is expected.
In this work we assume a conservative $\epsilon_{\rm EBL,1}(z)=1$ eV, where the peak of the stellar contribution is expected.
We checked that different values for the position of the EBL peak (in the energy range $0.5-2$ eV) affect the cascade normalization by less than $\lesssim 15 \%$, while leaving the spectral shape almost unaffected. 

In this approach, the electromagnetic cascade is assumed to develop instantaneously at the same redshift where the gamma rays are produced and subsequently absorbed by the EBL~\citep{Franceschini_EBL_erratum}. 
Such assumption is fully justified for the VHE photons that generate the cascade.
At a given redshift, the normalization of the cascade is computed assuming energy conservation, i.e. assuming that the total energy of gamma rays absorbed by the EBL is reprocessed in the electromagnetic cascade.

\section{Counting starburst sources}
\label{Sec_counting_prototype}


The standard definition of SBGs is based on properties of the IR-OPT spectral energy distribution or on the amount of IR emission above some threshold.
This definition is important from the observational point of view but it does not catch a crucial aspect of SBGs as sources of non--thermal radiation, namely their ability to effectively confine CRs.
Below we describe an attempt to retain this physical information by using the SFR while being clearly related to the infrared emission of the SBG.


\subsection{Distribution of SBNi}

\label{subsection_SF}

We adopt the SFRF approach described by~\citet{Gruppioni_2015_2} in the context of a study of the SFR using IR+UV data for a sample of Herschel sources \citep[see][]{Gruppioni_2013_Lumin_func} and subtracting the contamination of AGN as estimated by~\citet{Del_Vecchio_AGN}.

\begin{figure}
\centering
    \includegraphics[width=.21\textwidth]{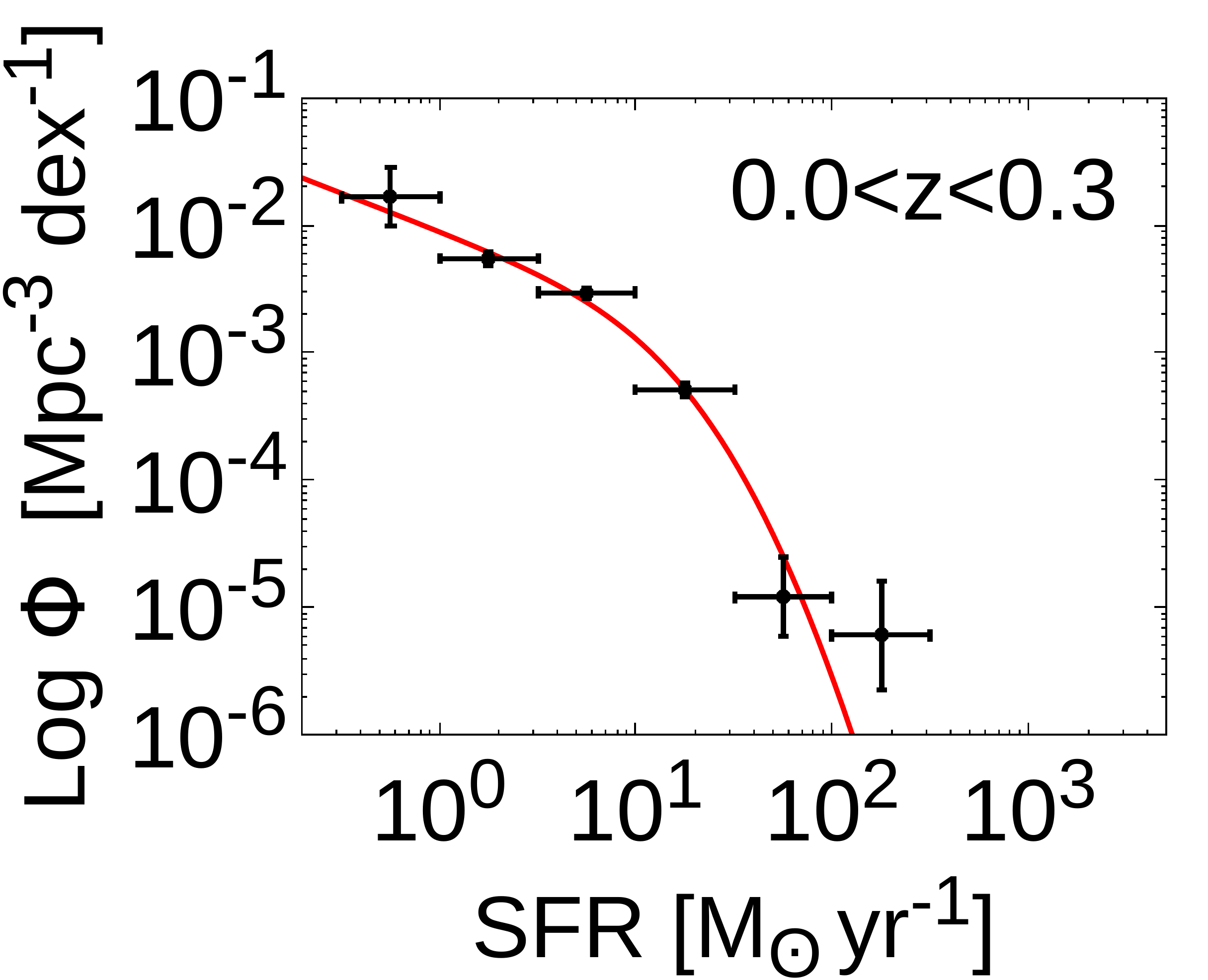}
    \includegraphics[width=.21\textwidth]{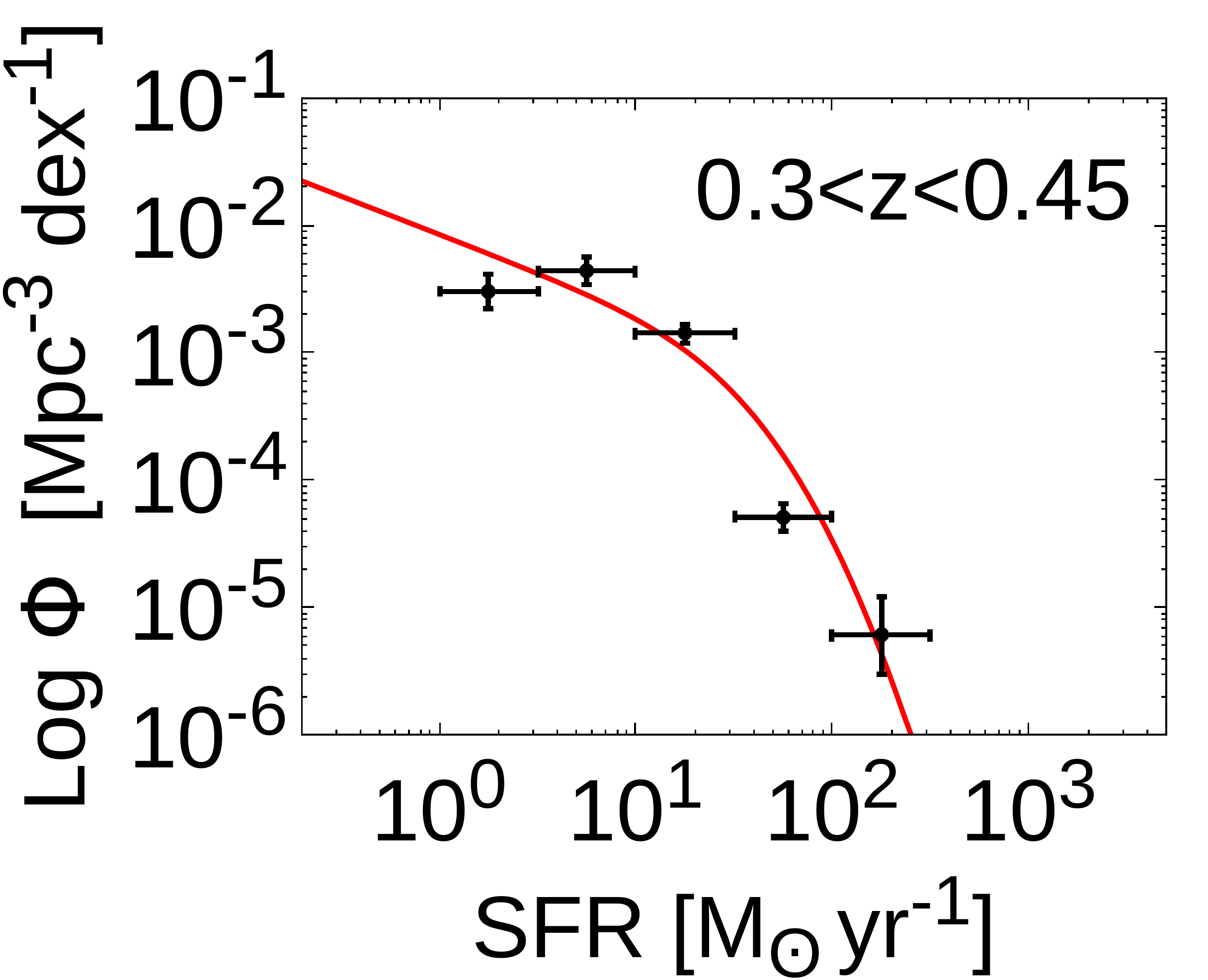}\\
    \includegraphics[width=.21\textwidth]{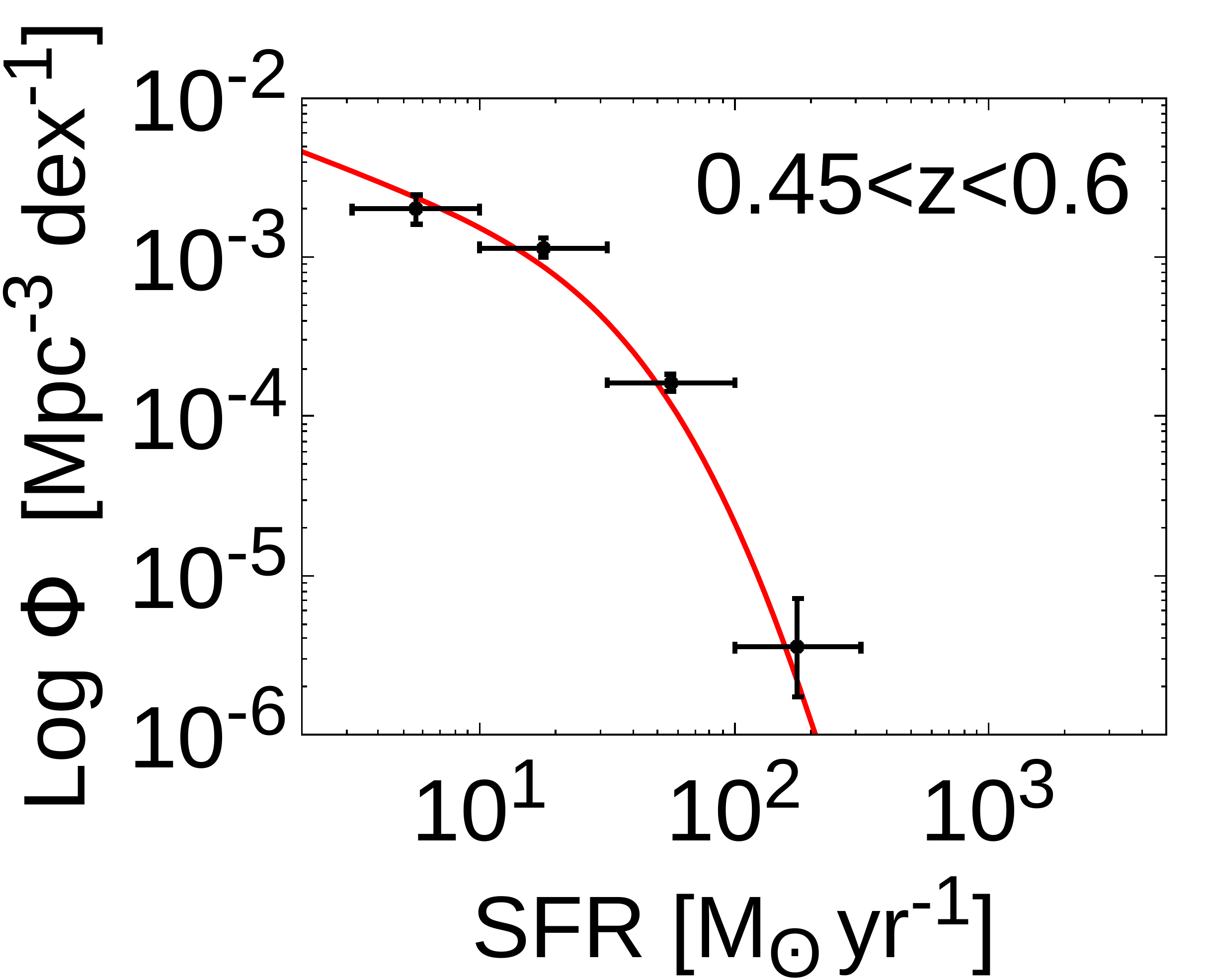} \quad
    \includegraphics[width=.21\textwidth]{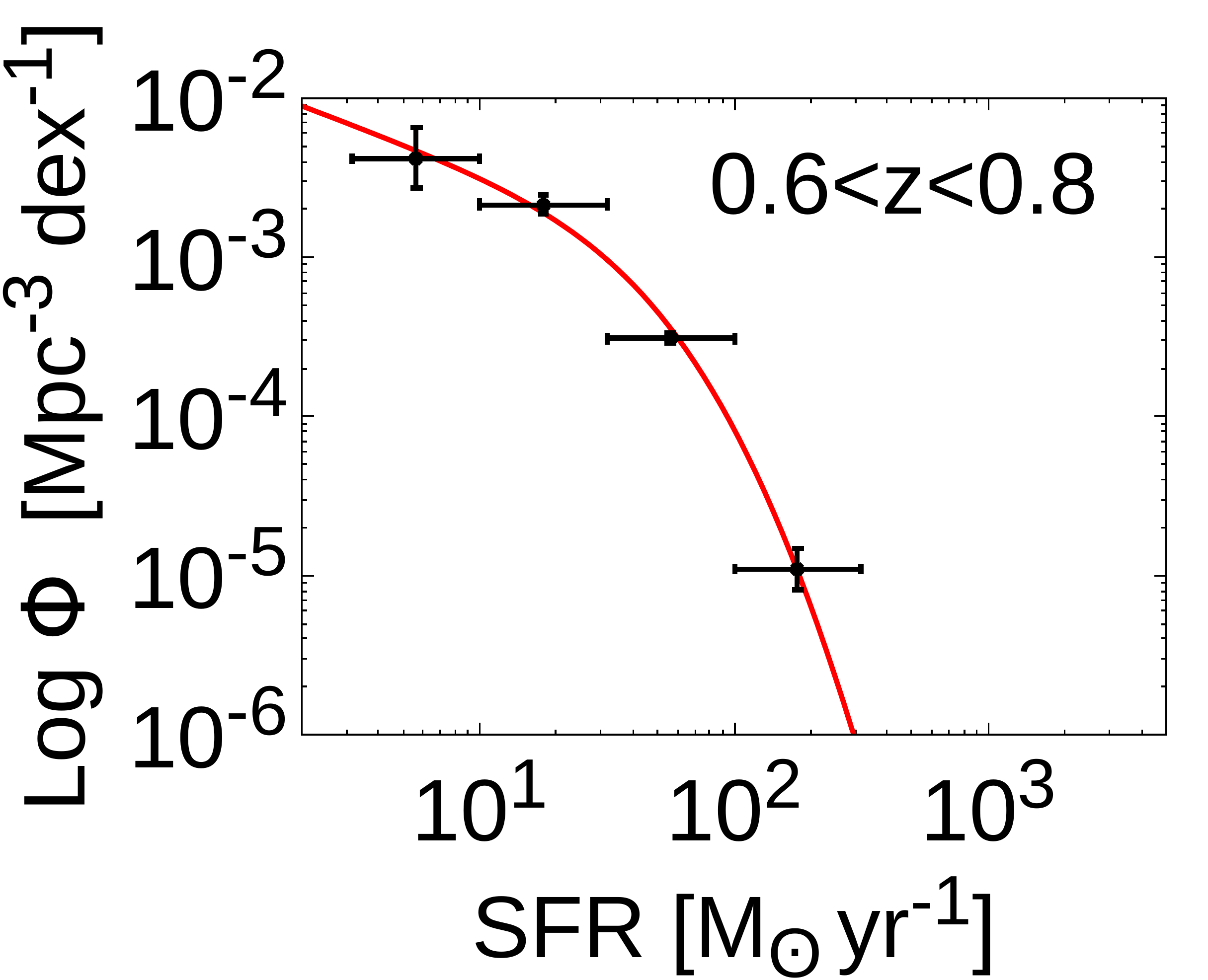}   \\
    \includegraphics[width=.21\textwidth]{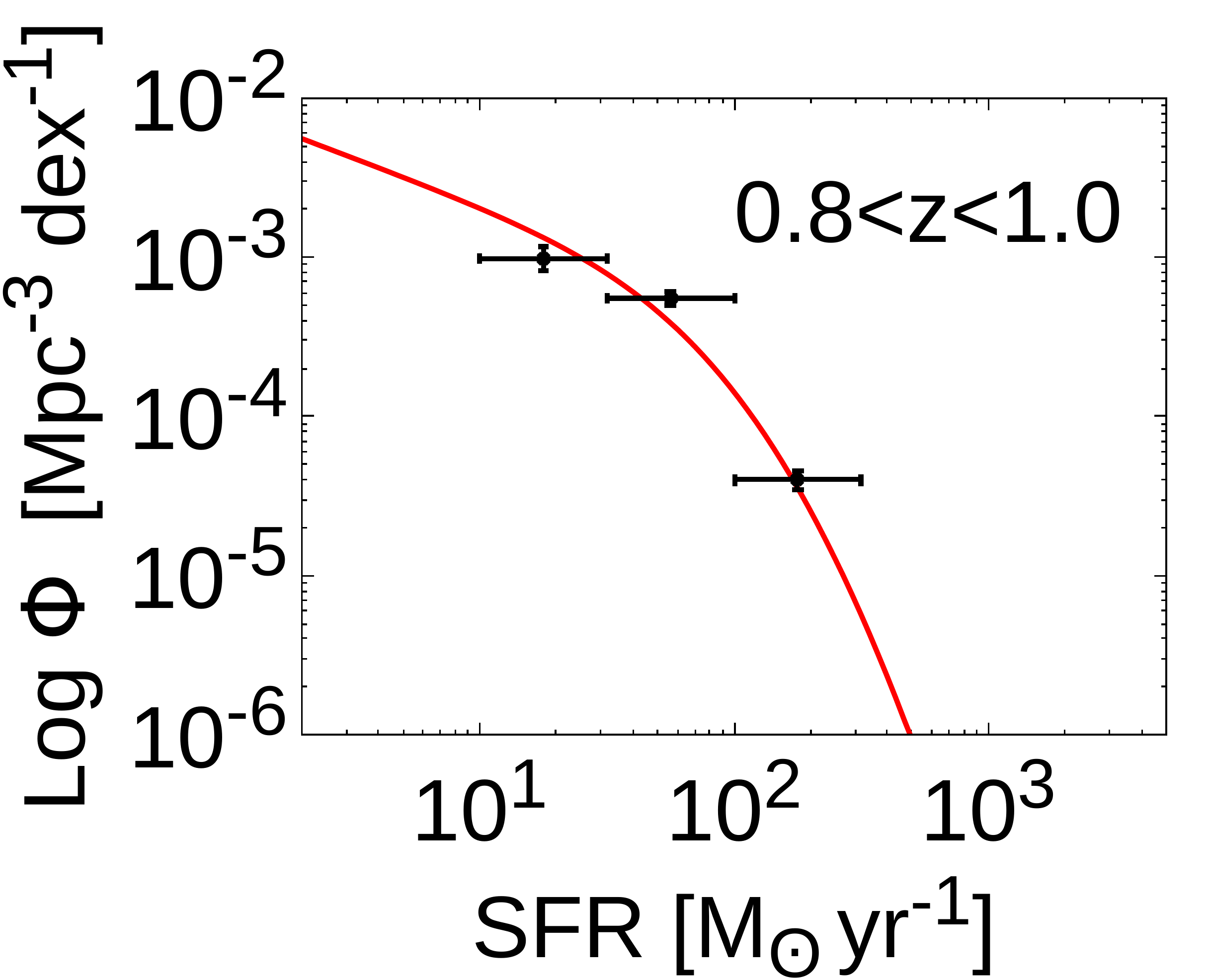} \quad
    \includegraphics[width=.21\textwidth]{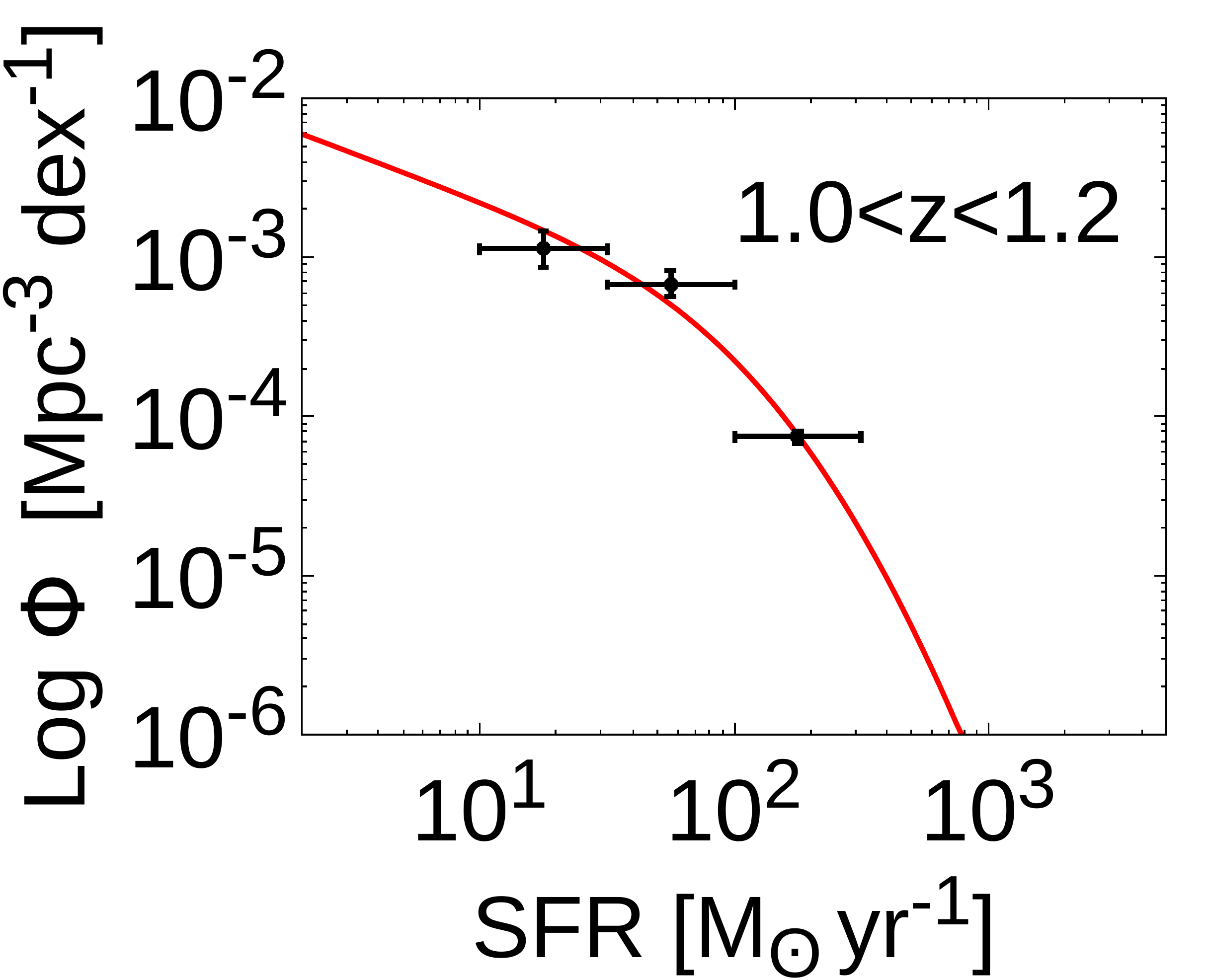} \\
    \includegraphics[width=.21\textwidth]{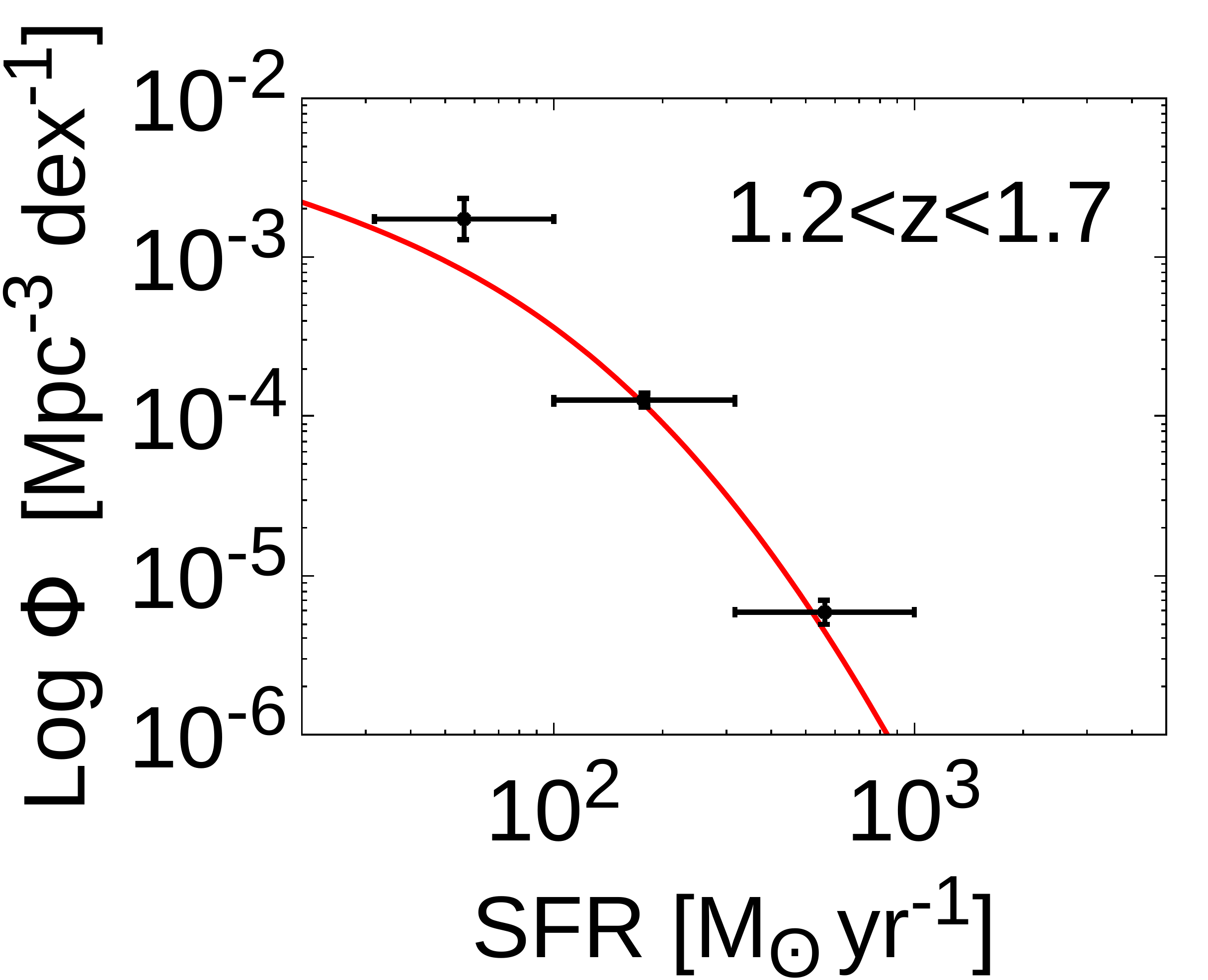} \quad
    \includegraphics[width=.21\textwidth]{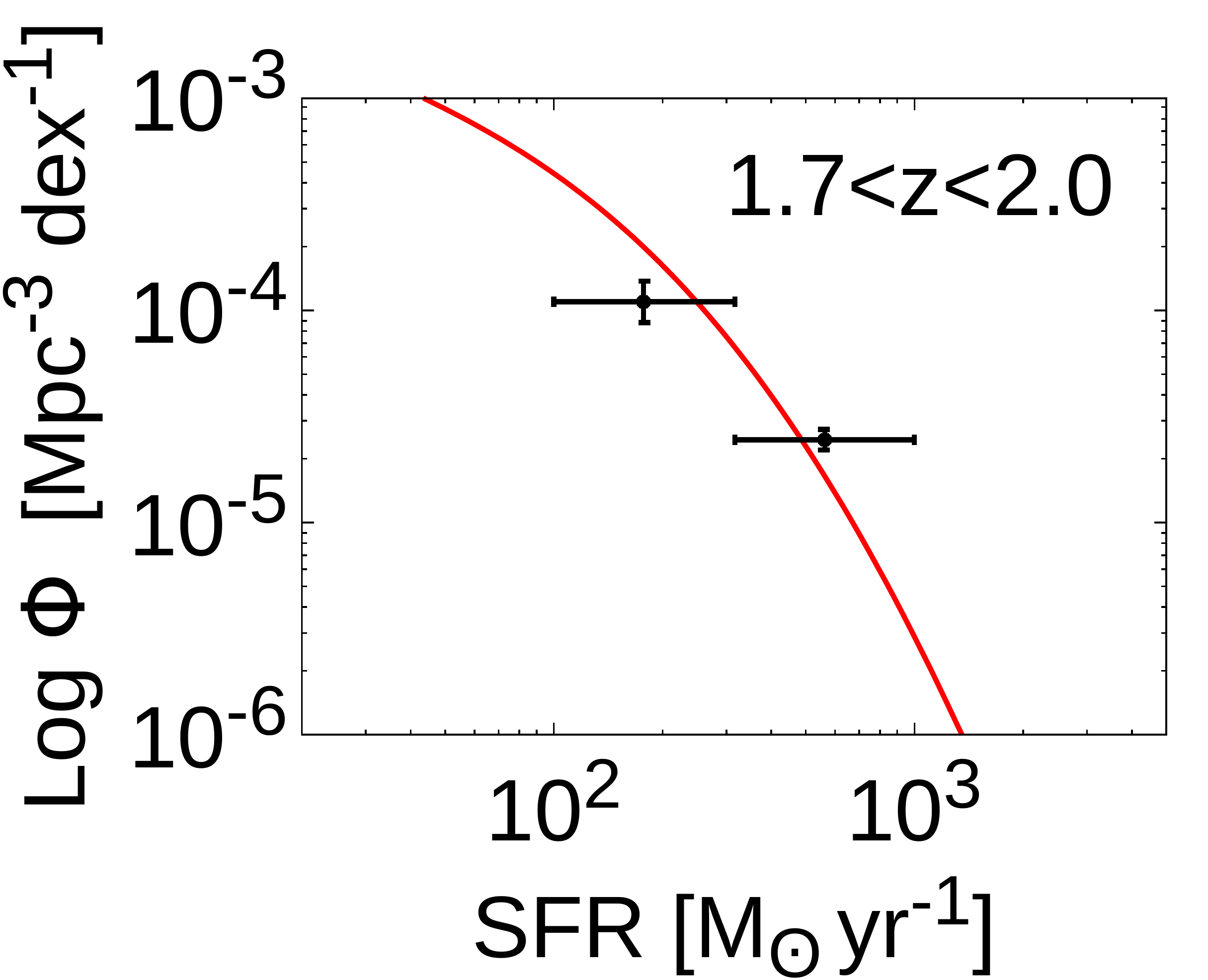}   \\
    \includegraphics[width=.21\textwidth]{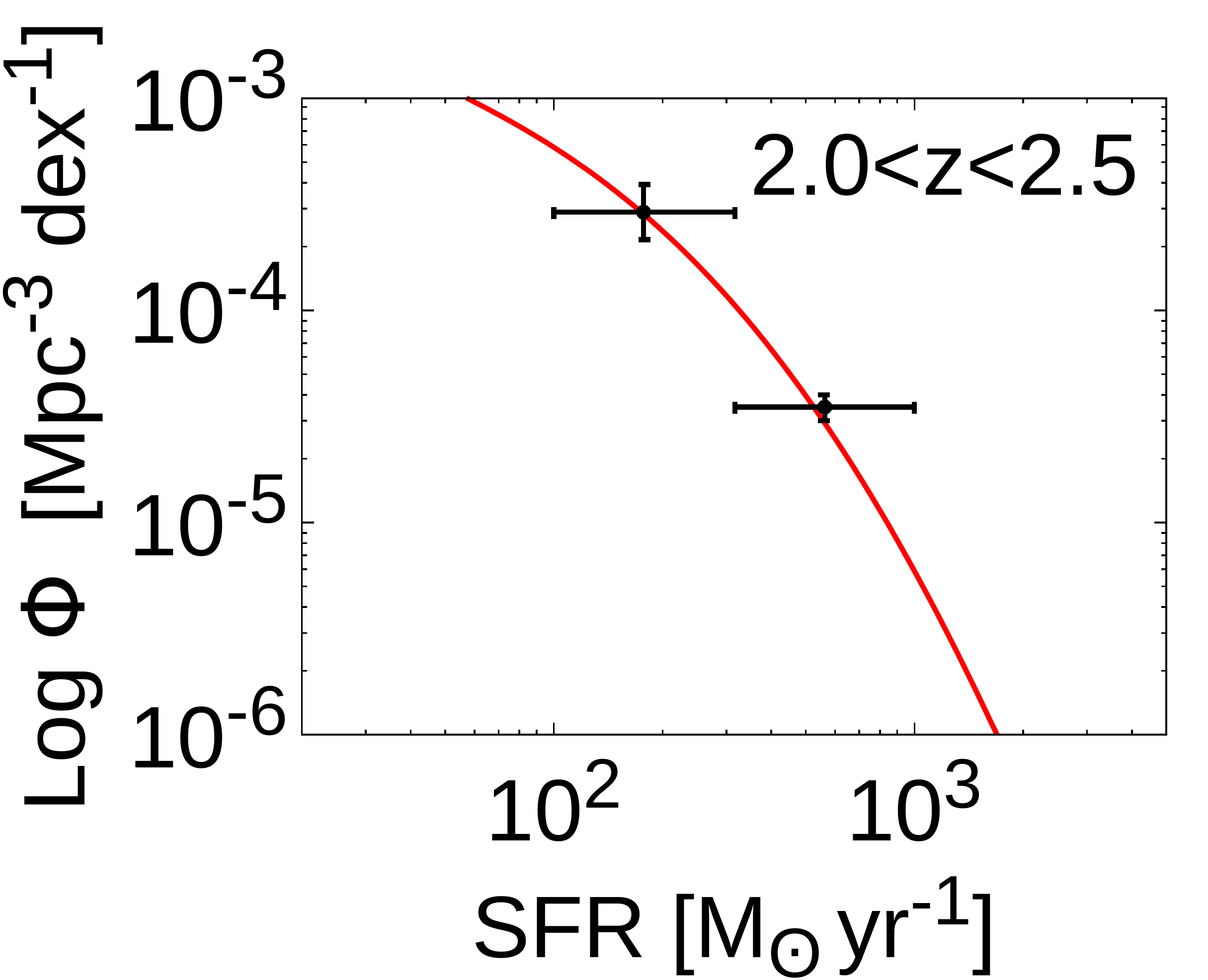} \quad
    \includegraphics[width=.21\textwidth]{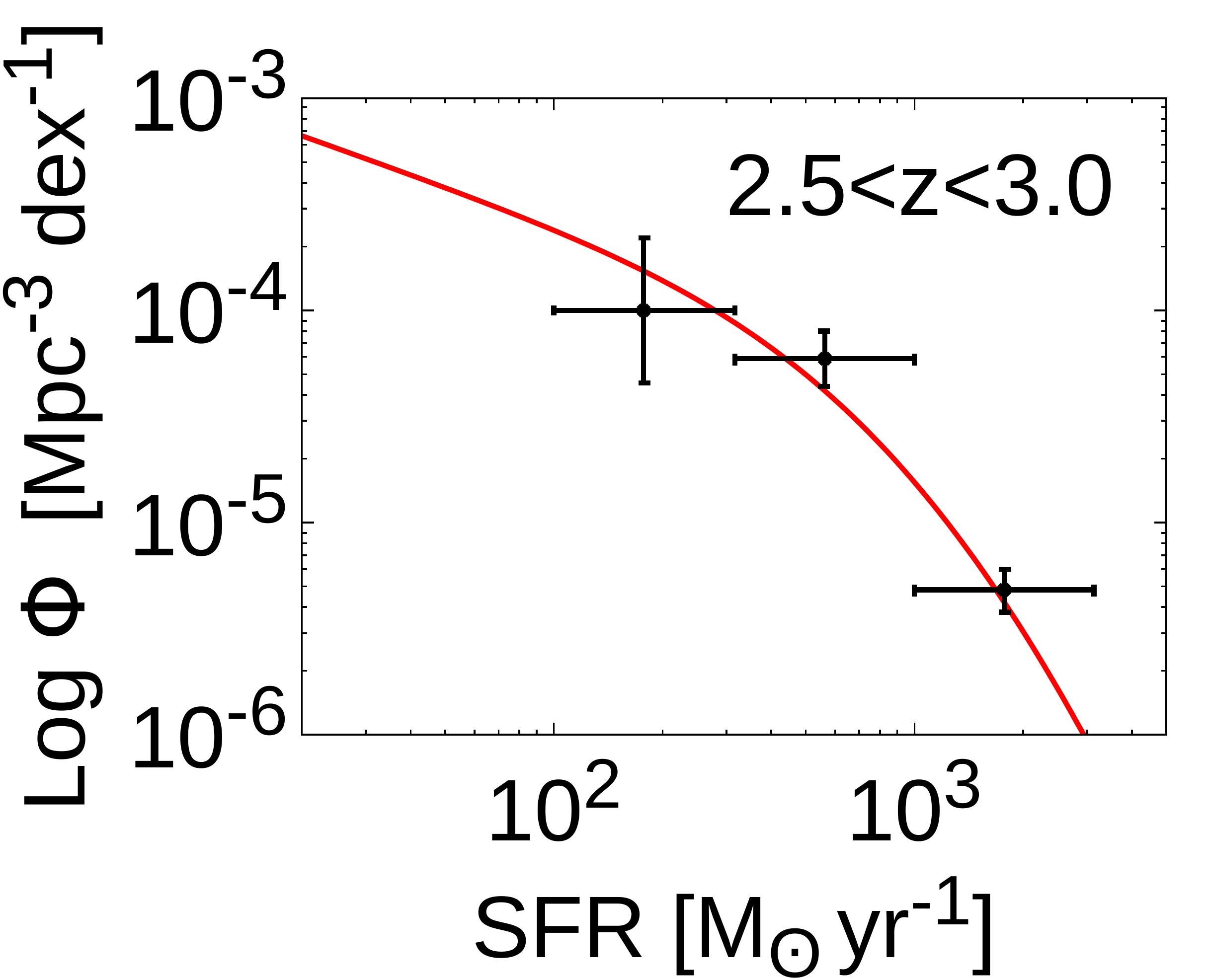} \\
    \includegraphics[width=.21\textwidth]{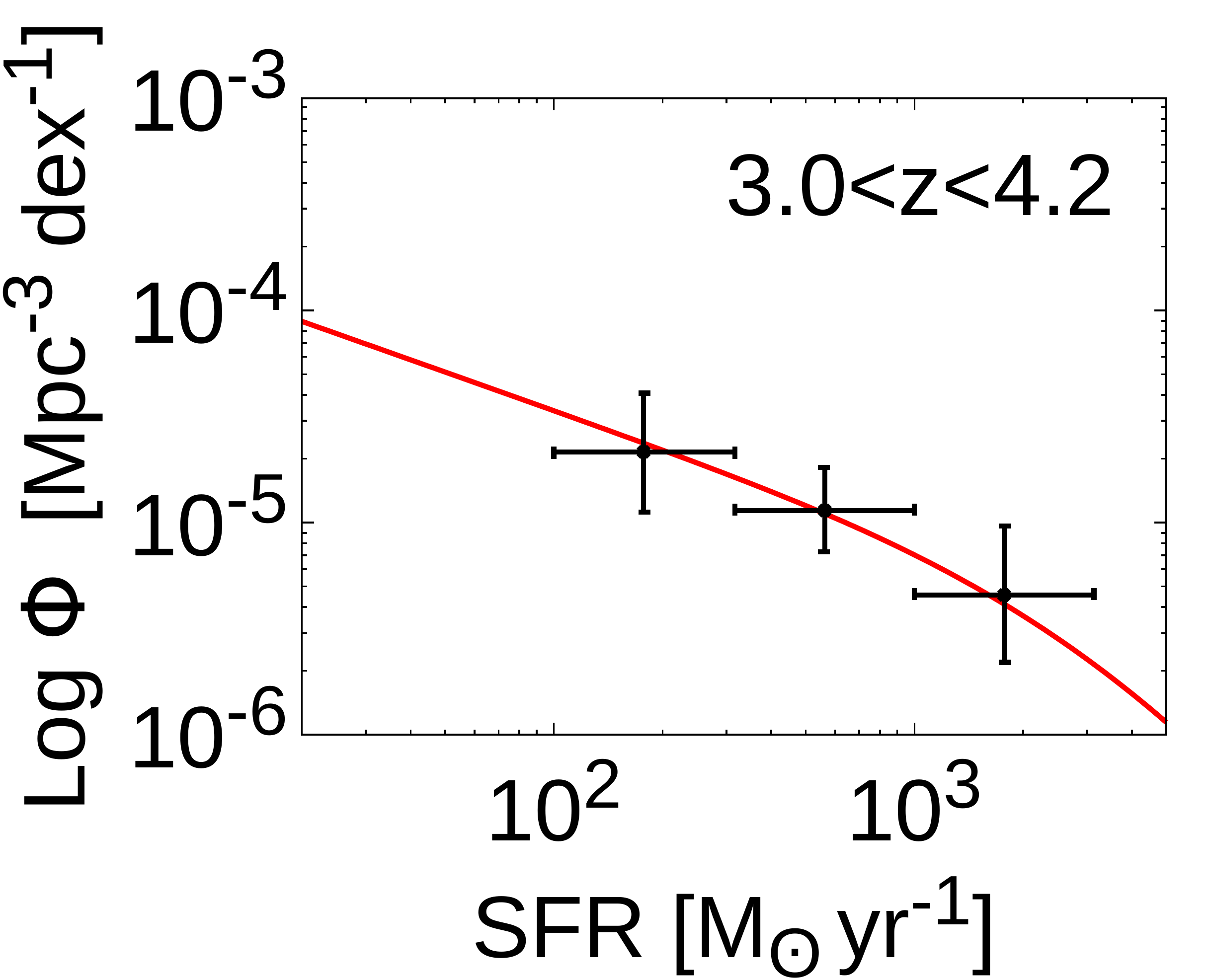} \quad
    \includegraphics[width=.21\textwidth]{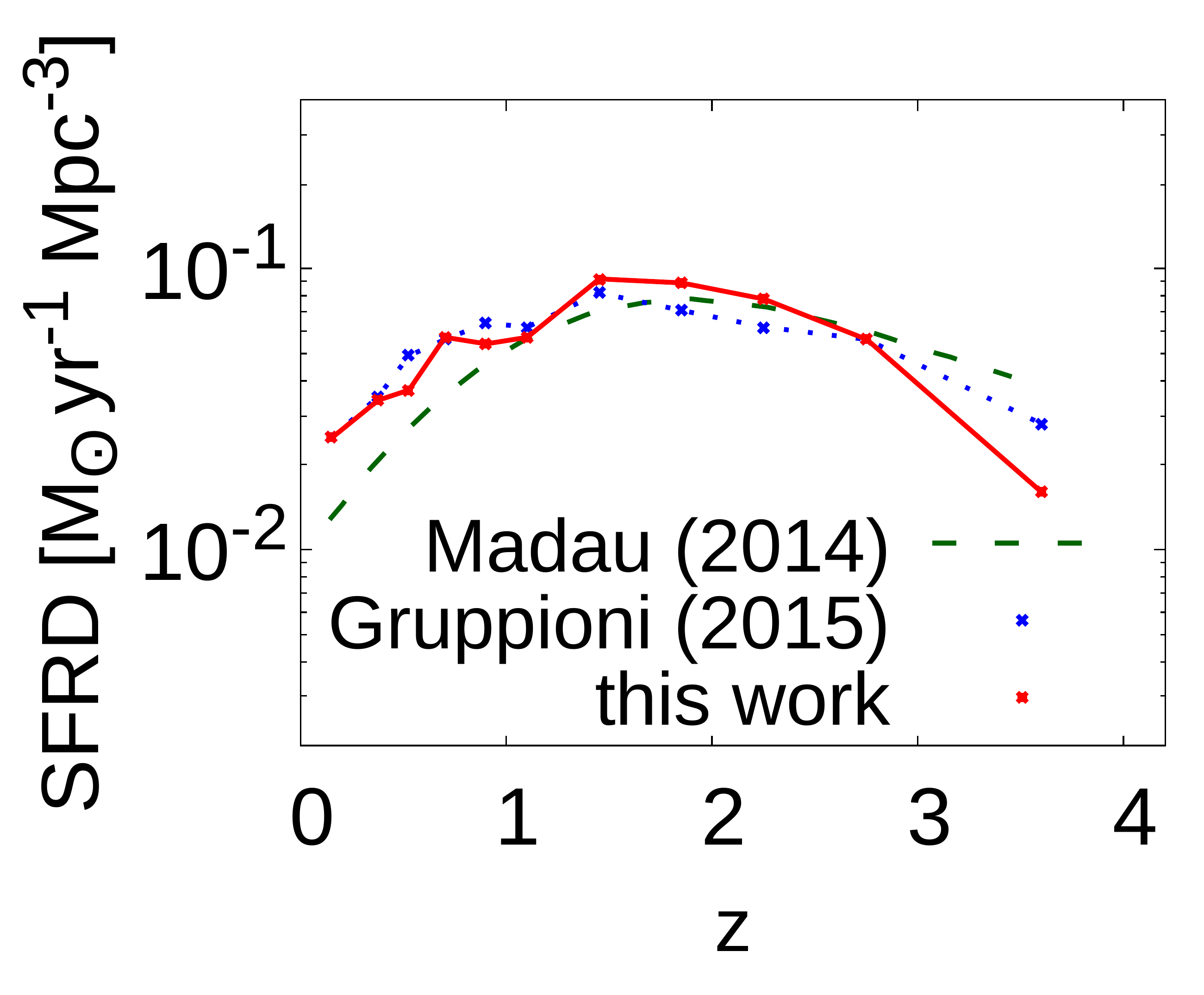}
  \caption{Best fit SFRF at each redshift (red curve) compared with data of~\citet{Gruppioni_2015_2} (black points). First row from left: $z=0.0-0.3$, $z=0.3-0.45$. Second row: $z=0.45-0.6$, $z=0.6-0.8$. Third row: $z=0.8-1.0$, $z=1.0-1.2$. Fourth row: $z=1.2-1.7$, $z=1.7-2.0$. Fifth row: $z=2.0-2.5$, $z=2.5-3.0$. Sixth row on the left $z=3.0-4.2$. The last plot in the bottom right corner shows our computed SFRD (thick red line) compared with the best fit values of \citet{Gruppioni_2015_2} (dotted blue line) and the best-fitting function of \citet{Madau:2014bja} (dashed green line).}
  \label{plot:SFRF_plot}
\end{figure}

For each redshift interval provided by \citet{Gruppioni_2015_2} we have derived our best fit, assuming the following modified Schechter function:
\begin{equation}
\label{SFRF_def}
    \Phi(\psi) \; d\log\psi= \Tilde{\Phi} \Big( \frac{\psi}{\Tilde{\psi}} \Big)^{1-\Tilde{\alpha}} \exp\Big[- \frac{1}{2 \Tilde{\sigma}^2} \log^2\Big( 1+\frac{\psi}{\Tilde{\psi}} \Big) \Big] \; d \log \psi,
\end{equation}
where $\psi$ is the SFR expressed in $\rm M_{\odot}$ yr$^{-1}$ and where
$\Tilde{\Phi}$, $\Tilde{\psi}$, $\Tilde{\alpha}$ and $\Tilde{\sigma}$ are redshift--dependent best fit parameters reported in Table~\ref{tab:parameters_table_2}. 
\begin{table}
\centering
\begin{tabular}{|c|c|c|c|c|}
\hline
 z   & $\Tilde{\Phi}$ $[10^{-3} M_{\odot}^{-1} dex^{-1}]$ & $\Tilde{\psi}$ & $\Tilde{\alpha}$ & $\Tilde{\sigma}^2$ $[10^{-1}]$ \\ \hline \hline
 $0.0-0.3$  & $2.8$ & $7$ & $1.6$ & $1.32$   \\ \hline
 $0.3-0.45$ & $1.5$ & $18$ & $1.6$ & $1.2$   \\ \hline
 $0.45-0.6$ &  $1.2$ & $27$ & $1.6$ & $0.85$   \\ \hline
 $0.6-0.8$ &  $1.5$ & $34$ & $1.6$ & $0.8$   \\ \hline
 $0.8-1.0$ &  $1.2$ & $32$ & $1.6$ & $1.5$   \\ \hline
 $1.0-1.2$ & $1.05$ & $36$ & $1.6$ & $1.8$   \\ \hline
 $1.2-1.7$ &  $1.7$ & $37$ & $1.6$ & $1.7$   \\ \hline
 $1.7-2.0$ &  $0.9$ & $65$ & $1.6$ & $1.8$   \\ \hline
 $2.0-2.5$ & $0.35$ & $170$ & $1.6$ & $1.2$   \\ \hline
 $2.5-3.0$ &  $0.15$ & $240$ & $1.6$ & $1.8$   \\ \hline
 $3.0-4.2$ &  $0.0145$ & $550$ & $1.6$ & $3.5$   \\ \hline
\end{tabular}
\caption{Fit parameters of the SFRF  for each considered redshift interval.}
\label{tab:parameters_table_2}
\end{table}
In Fig.~\ref{plot:SFRF_plot} we show, for each redshift interval, the best fit SFRF corresponding to the list of values reported in Table~\ref{tab:parameters_table_2} compared with data of~\citet{Gruppioni_2015_2}. Moreover in the bottom right panel we compare the redshift behaviour of the inferred star formation rate density (SFRD), namely the integral of the SFRF weighted by $\psi$, with what was obtained by \citet{Gruppioni_2015_2} and \citet{Madau:2014bja}, showing that our result are compatible with the latter ones.
In Table~\ref{tab:parameters_table} we provide for each redshift interval the value of the SFRD computed assuming
$\log \psi_{\rm min}=-1.5$ as minimum SFR, as in~\citet{Gruppioni_2015_2}. 
We also verified a posteriori that our obtained SFRD is compatible within $1 \; \sigma$ uncertainty with the results obtained by \citet{Gruppioni_2015_2} (shown for comparison in the same table).
\begin{table}
\centering
\begin{tabular}{|c|c|c|}
\hline
 z  &  SFRD (this work) & SFRD \citep[see][]{Gruppioni_2015_2}  \\ \hline \hline
 $0.0-0.3$ & $0.025$ & $0.025 \pm 0.005$  \\ \hline
 $0.3-0.45$ & $0.034$ & $0.035 \pm 0.010$ \\ \hline
 $0.45-0.6$ & $0.037$ & $0.049 \pm 0.014$ \\ \hline
 $0.6-0.8$ & $0.057$ & $0.056 \pm 0.013$  \\ \hline
 $0.8-1.0$ & $0.054$ & $0.064 \pm 0.016$  \\ \hline
 $1.0-1.2$ & $0.057$ & $0.062 \pm 0.014$  \\ \hline
 $1.2-1.7$ & $0.092$ & $0.082 \pm 0.021$  \\ \hline
 $1.7-2.0$ & $0.089$ & $0.071 \pm 0.019$  \\ \hline
 $2.0-2.5$ & $0.078$ & $0.062 \pm 0.021$  \\ \hline
 $2.5-3.0$ & $0.056$ & $0.056 \pm 0.020$  \\ \hline
 $3.0-4.2$ & $0.016$ & $0.028 \pm 0.012$  \\ \hline
\end{tabular}
\caption{SFRD in units of $\rm (M_{\odot} \, yr^{-1} \, Mpc^{-3})$ for each considered redshift interval. The SFRD is obtained integrating $\Phi(\psi)$ for $\log_{10}(\psi) \geq -1.5$.}
\label{tab:parameters_table}
\end{table}

For our calculations, it is important to clearly define the properties that a galaxy should possess to be considered as an SBN.
Many definitions of SBNi are present in the literature, based e.g. on  gas consumption, star formation rate, gas surface density, or star formation rate density \citep[see e.g.][for a detailed review]{Kennicutt2012}.
Here we adopt a somewhat different definition, more closely connected with the non--thermal activity of the SB region and based on its ability to effectively confine CRs inside the nuclear region. 

Focusing on the requirement of CR confinement, we define an SBN as a source where the timescale for losses is shorter than the dominant escape time. As discussed in P18, for typical conditions, the escape time is dominated by advection, hence the condition is simply
\begin{equation}
    \tau_{\rm loss} \leq \tau_{\rm adv}.
    \label{Req_n}
\end{equation}
Here $\tau_{\rm loss} \approx 1/n_{\rm ISM}\sigma_{pp} c \eta$, where $\eta \approx 0.5$ is the inelasticity for pp collisions, and $\tau_{\rm adv}\approx R/v_{\rm wind}$ is the advection time. We introduce the gas surface density as $\Sigma_{gas} = n_{\rm ISM} m_p R$, so that Eq. \eqref{Req_n} leads to the following condition: 
\begin{equation}
\label{GasSD}
    \Sigma_{\rm gas} \geq \Sigma_{\rm gas}^{*} \approx 1068 \; \left[\frac{v_{\rm wind}}{10^3 \, \text{km/s}} \right] \; \frac{M_{\odot}}{pc^2},
\end{equation}
where we have assumed $\sigma_{pp}\approx 50$ mb. 
The obtained critical gas surface density $\Sigma^*_{\rm gas}$ is compatible with  $\Sigma_{\rm SB} \gtrsim [1\div 3] \times 10^2$ M$_{\odot}$pc$^{-2}$ expected for high star forming regions and a factor $\sim 4$ larger than what is inferred for the CMZ of the Milky Way \citep[][]{Kennicutt2012}. 

Our \textit{confinement requirement} expressed by Eq. \eqref{Req_n} shows its most natural implication in terms of the injection spectra of hadronic byproducts like gamma rays and neutrinos which become only a function of the injection spectrum of their parent protons. 
Such injection spectrum can be well approximated by: 
\begin{equation}
\label{nuegamma}
    q_{\gamma,\nu}(E) = [n_{\rm ISM} \sigma_{pp} c] f_p(E/\kappa_{\gamma,\nu})/\kappa_{\gamma,\nu},
\end{equation}
where $\kappa_{\gamma,\nu}$ is the energy transferred from a parent proton to a secondary gamma--ray or neutrino. 
The proton distribution function $f_p$ is the solution of the transport equation (see Eq. \eqref{Transport_solution}) where $\tau_{\rm diff}$ can be neglected. 
The asymptotic expression of $f_{p}$ is governed by the minimum timescale: 
\begin{equation}
    f_p(p) \propto 
    \begin{cases}
    q_p(p) \; \tau_{\rm loss} & \tau_{\rm loss} \ll \tau_{\rm adv} \\
    q_p(p) \; \tau_{\rm adv} & \tau_{\rm loss} \gg \tau_{\rm adv}
    \end{cases}.
    \label{Asymptotic_1}
\end{equation}
Substituting the asymptotic expressions of Eq. \eqref{Asymptotic_1} in Eq. \eqref{nuegamma}, one obtains: 
\begin{equation}
\label{inj_behav}
    q_{\gamma,\nu}(E) \propto 
    \begin{cases}
    q(p) & \tau_{\rm loss} \ll \tau_{\rm adv} \\
    [n_{\rm ISM} \sigma_{pp} c] q_p(p) R/ v_{\rm wind} & \tau_{\rm loss} \gg \tau_{\rm adv}
    \end{cases}.
\end{equation}
In the calorimetric scenario, i.e. when $\tau_{\rm loss} \ll \tau_{\rm adv}$, the production of gamma rays and neutrinos is proportional to the injection of primary protons: it therefore depends on the rate of SNe (see Eq.~\eqref{SNe inj}).
On the other hand, in the advection dominated scenario, i.e. when $\tau_{\rm adv} \ll \tau_{\rm loss}$, the production of gamma rays and neutrinos is also proportional to the gas surface density $\Sigma_{gas}\propto n_{\rm ISM}R$ and inversely proportional to the wind speed $v_{\rm wind}$, typically inferred from observations in the range $\sim 10^2 \div 10^3$ km/s. 

From the minimum value of the surface density of gas $\Sigma^*_{gas}$ expressed in Eq. \eqref{GasSD} it is possible to infer the associated surface density of SFR adopting the Kennicutt relation \citep{Kennicutt:1997Relation}:
\begin{align}
    \label{Kennicutt relation}
    \frac{\Sigma_{\rm SFR}^*}{M_{\odot} \text{yr}^{-1} \text{kpc}^{-2}} & = (2.5 \pm 0.7) \times 10^{-4} \left[ \frac{\Sigma_{\rm gas}^*}{1 \; M_{\odot}\text{pc}^{-2}} \right]^{1.4 \pm 0.15} \\
    & = 4.35^{+11.49}_{-3.25} \nonumber.
\end{align}
The obtained value is fully compatible with what is expected for circumnuclear regions in star forming galaxies \citep[see e.g.][]{Kennicutt:1998zb} and allows to compute the associated value of SFR as:
\begin{equation}
\label{min_SFR}
    \psi^* = \Sigma_{\rm SFR}^* \pi R^2 \approx 0.9^{+2.2}_{-0.7} \; \Big[\frac{R}{0.25 \; \text{kpc}}\Big]^2 \; \rm M_{\odot} \rm \text{yr}^{-1}.
\end{equation}
Such a definition of a minimum value of the SFR required for an SBN to be an efficient calorimeter, allows to perform the number counting of galaxies at every redshift by integrating the SFRF $\Phi(\psi)$ for $\psi \geq \psi^*$. 
Hereafter we assume that the star--forming activity of all galaxies with $\psi \geq \psi^*$ is localized in the SBN.

\subsection{SBN prototype}
\label{Subsection_prototype}

In order to compute the diffuse flux, we rely on the SFRF approach and we adopt a prototype SBG following the model described in  Sec. \ref{Sezione1}. 
In the following we will adopt M82 as a prototype, being one of the best studied nearby galaxies possessing a nuclear region in a starburst phase. 

\begin{table}
\centering
\begin{tabular}{|c|c|}
\hline
 parameter  &  value  \\ \hline \hline
 $p_{p, \text{max}}$ & $10^2$ PeV \\ \hline 
 $\alpha$ &  $4.2$ \\ \hline  \hline
 $R$ &  $0.25$ kpc \\ \hline
 $D_L$ & $3.9$ Mpc  \\ \hline
 $\xi_{\rm CR}$ & $0.1$ \\ \hline  \hline
 $\mathcal{R}_{\rm SN}$ &  $0.06$ yr$^{-1}$  \\ \hline
 $B$ &  $200$ $\mu$G  \\ \hline
 $n_{\rm ISM}$ &  $100$ cm$^{-3}$  \\ \hline
 $v_{\rm wind}$ &  $700$ km/s  \\ \hline
 $U_{\rm rad}$ &  $2500$ eV/cm$^3$  \\ \hline
\end{tabular}
\caption{Parameters for the starburst M82-like prototype. The first five parameters are fixed: maximum momentum, injection slope, SBN radius, luminosity distance and CR acceleration efficiency. The last five parameters are kept free in the fitting procedure: SN rate, magnetic field, ISM density, wind speed and radiation energy density.}
\label{tab:M82-like-fit}
\end{table}

The size (radius) of the prototypical SBN is set to $250 \, \rm pc$, an average value for these circumnuclear regions.
The spectrum of CRs at injection is chosen to be $\propto p^{-4.2}$, consistent with what is inferred for observed starbursts (see also dedicated calculations in P18).
The other parameters, listed in Tab. \ref{tab:M82-like-fit}, have been obtained looking for the best fit to M82 data. 
We found a good agreement between our results and the values quoted in the literature \citep{FenechM82,Yoast-Hull_M82_2013}.
\begin{figure}
\centering
\includegraphics[width=0.45\textwidth]{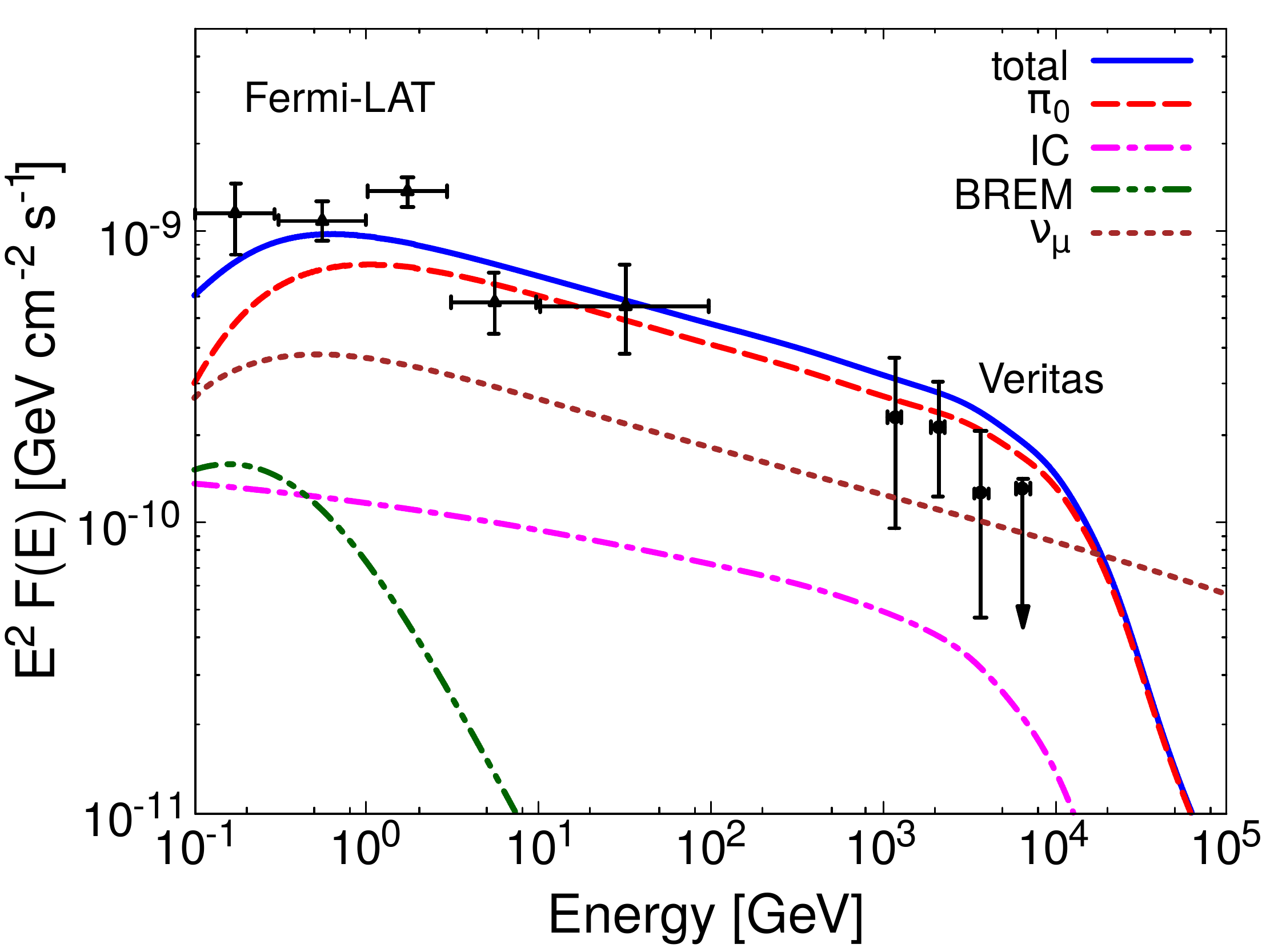}  \caption{\label{fig:M82like} High--energy gamma rays (blue thick line) and neutrino (brown dotted line) spectra of our prototype SBN compared with M82 data from Fermi--LAT and Veritas \citep{3FGL,Veritas_M82}. Inverse Compton (magenta dot-dashed), bremsstrahlung (green dot-dot-dashed) and $\pi_0$ decay (red dashed) components are shown.}
\end{figure}
In Fig. \ref{fig:M82like} we show the multiwavelength electromagnetic  and neutrino spectra of our M82--like prototype.

The SN rate is first obtained by fitting the multi--wavelength emission of M82. Second, the SFR is obtained assuming that one supernova is exploding every 100 $M_{\odot}$ converted in new stars. 
Our fit is compatible with a SN rate in the range [0.04 -0.08] $\rm yr^{-1}$, therefore we adopt as reference value $\mathcal{R}_{\rm SN}^{\rm M82}= 0.06 \, \rm yr^{-1}$ and consequently $\psi_{\rm M82}= 6 \, \rm M_{\odot} \, yr^{-1}$.

If the \textit{confinement requirement} is satisfied, Eq. \eqref{inj_behav} shows that the gamma--ray and neutrino luminosities scale linearly with the primary proton injection rate, which in turn is proportional to the rate of supernovae $\mathcal{R}_{\rm SN}$.
We therefore can write the luminosity for a general SBN as linear function of the SFR:
\begin{equation}
\label{SB_prototype}
    f^{\rm SBN}_{\gamma,\nu}(E,\psi) = \Big( \frac{\psi}{\psi_{\rm M82}} \Big) f^{M82}_{\gamma,\nu}(E), 
\end{equation}
where $f^{\rm SBN}_{\gamma,\nu}$ and $f^{M82}_{\gamma,\nu}$ are respectively the flux densities of gamma rays and neutrinos for a generic SBN and for our prototype. 
Eq.~\eqref{SB_prototype}, though well motivated~\citep[see also IR-$\gamma$ scaling][where the IR luminosity is also a SFR tracer]{Ackermann_Fermi_2012}, might overestimate the gamma--ray flux at $E\gtrsim$ TeV. 
This is due to the fact that for SBNi with a higher SFR a stronger gamma--gamma absorption is expected to occur inside the source due to the intense low energy photon fields produced by stars and dust. Such unabsorbed radiation then leads to a larger energy budget for the electromagnetic cascade. By assuming the FIR--OPT background of M82 for each SBN, we do not take into account this effect.

The validity of the linear dependence expressed by Eq. ~\eqref{SB_prototype} can be tested by comparing the cases of M82 and Arp220. The gamma--ray flux of the prototypical M82 as due to $\pi_0$ decays is $L_{\pi_0} \approx 1.82 \times 10^{40}$ erg s$^{-1}$ to be compared with $L_{\pi_0} \approx 1.29 \times 10^{42}$ erg s$^{-1}$ for Arp220. 
The corresponding SFR for Arp220 is of the order of $400 \, \rm M_{\odot} \, yr^{-1}$, which is compatible with results in the literature which range between 260 and $\sim 580 \, \rm M_{\odot} \, yr^{-1}$  \citep[see, e.g.][]{Groves:2007py}.

\section{Results}
\label{Section_results}

Relying on the approach described in Sec.~\ref{Sezione1} and Sec.~ \ref{Sec_counting_prototype}, we compute the diffuse gamma--ray and neutrino spectra integrating the emitted flux over the cosmological history:
\begin{align}
    \Phi_{\gamma,\nu}(E)= &  \frac{1}{4 \pi} \int d \Omega \int_0^{4.2} dz \; \frac{dV_{\rm C}{ (z) }}{dz \, d \Omega} \, \times \nonumber \\   
    & \int_{\psi^{*}} d \log \psi \; \Phi_{\rm SFR}(\psi,{ z}) \; {[1+z]^2} f_{\gamma,\nu}(E{ [1+z]},\psi).
    \label{SFRF}
\end{align}
Here $dV_{\rm C}= c D_{\rm C}^2(z)/[E(z) \, H_0] \, dz \, d \Omega $ is the comoving volume element per redshift interval $dz$ and solid angle $d \Omega$, where in a flat space $D_{\rm C}(z)= D_{\rm L}(z)/(1+z)$ and $E(z)= \sqrt{\Omega_{\rm M} (1+z)^3 + \Omega_{\Lambda}}$ , and the $f_{\gamma,\nu}(E,\psi)$ is the flux density and accounts for the dependence on the SFR given by Eq.~\eqref{SB_prototype}.

We define as \textit{benchmark case} the diffuse gamma--ray and neutrino flux computed adopting the prototypical SBN described in Sec.~\ref{Subsection_prototype}, where we have assumed injection slope $\alpha=4.2$, maximum momentum $p_{p, \rm max}= 100 \, \rm PeV$ and SFR $\psi_{\rm M82}= 6.0 \, M_{\odot} \, yr^{-1}$.
In the following, we discuss this benchmark case and the impact of changing the value of these three parameters.

As discussed in the introduction, the main constraint in this analysis comes from the blazar contamination of the gamma--ray flux observed by Fermi--LAT in the energy band $50 \, \rm GeV - 2 \, TeV$ \citep[][]{Fermi-LAT>50,Lisanti_2016}. We consider, as firm upper limit for such flux, $\sim 40 \%$ of the total EGB observed in that energy band obtained requiring $1 \, \sigma$ compatibility with \citet{Lisanti_2016} (this corresponds to $ \approx 9.6 \times 10^{-10} \, \rm ph \, cm^{-2} \, s^{-1} \, sr^{-1}$). 

\begin{figure}
\centering
\includegraphics[width=0.45\textwidth]{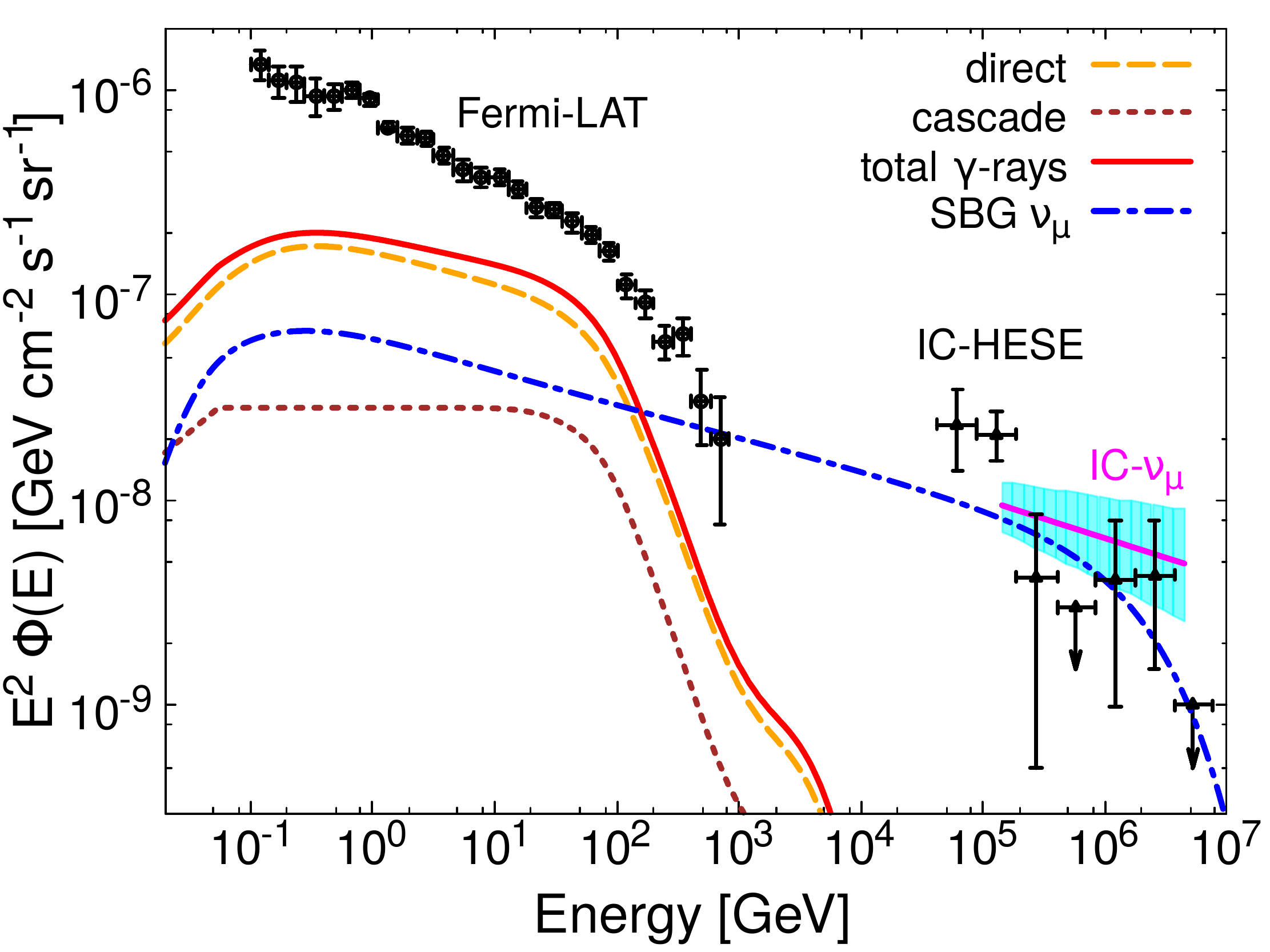}  \caption{\label{fig:Diff_ref} Diffuse gamma--ray (thick red) and single-flavor neutrino (dot-dashed blue) fluxes of starburst origin computed with our benchmark case and compared with Fermi-LAT EGB \citep[see][]{Fermi-LAT<820}, and neutrino HESE and through-going muon samples \citep[see][]{LASTHESE_taboada_ignacio_2018_1286919,IceCube_muons} (shaded band). 
We also show the individual contribution due to direct gamma rays (orange dashed) and the electromagnetic cascade (brown dotted).}
\end{figure}

In Fig.~\ref{fig:Diff_ref} the results for our benchmark case are shown, for both gamma--rays and neutrinos. For the gamma--ray flux we specify the direct and cascade contribution. The neutrino flux refers to a single flavor. The total gamma--ray flux is found to be well below the total EGB in agreement with \cite{Fermi-LAT<820}. The gamma--ray flux at energies $>50 \, \rm GeV$ is close to saturating the diffuse background but still compatible with the upper limit (more specifically we obtain $\approx 9.2 \times 10^{-10} \, \rm ph \, cm^{-2} \, s^{-1} \, sr^{-1}$). This case clearly illustrates the fact that both SBNi and blazars may both be significant contributors to the gamma--ray flux above 50 GeV.  

Our computed single flavor neutrino flux is about a factor $\sim 2$ below the measured HESE for energy below $\sim 200$ TeV, whereas it is compatible with neutrino data at higher energies. The discrepancy between our prediction and the observed flux is however within $\sim 2\sigma$. A very good agreement between our computed neutrino flux and the through-going muon data sample~\citep[][]{IceCube_muons} (violet line with the $1\sigma$ compatibility region shown as an azure band) is found. 

\begin{figure}
\centering
\includegraphics[width=0.44\textwidth]{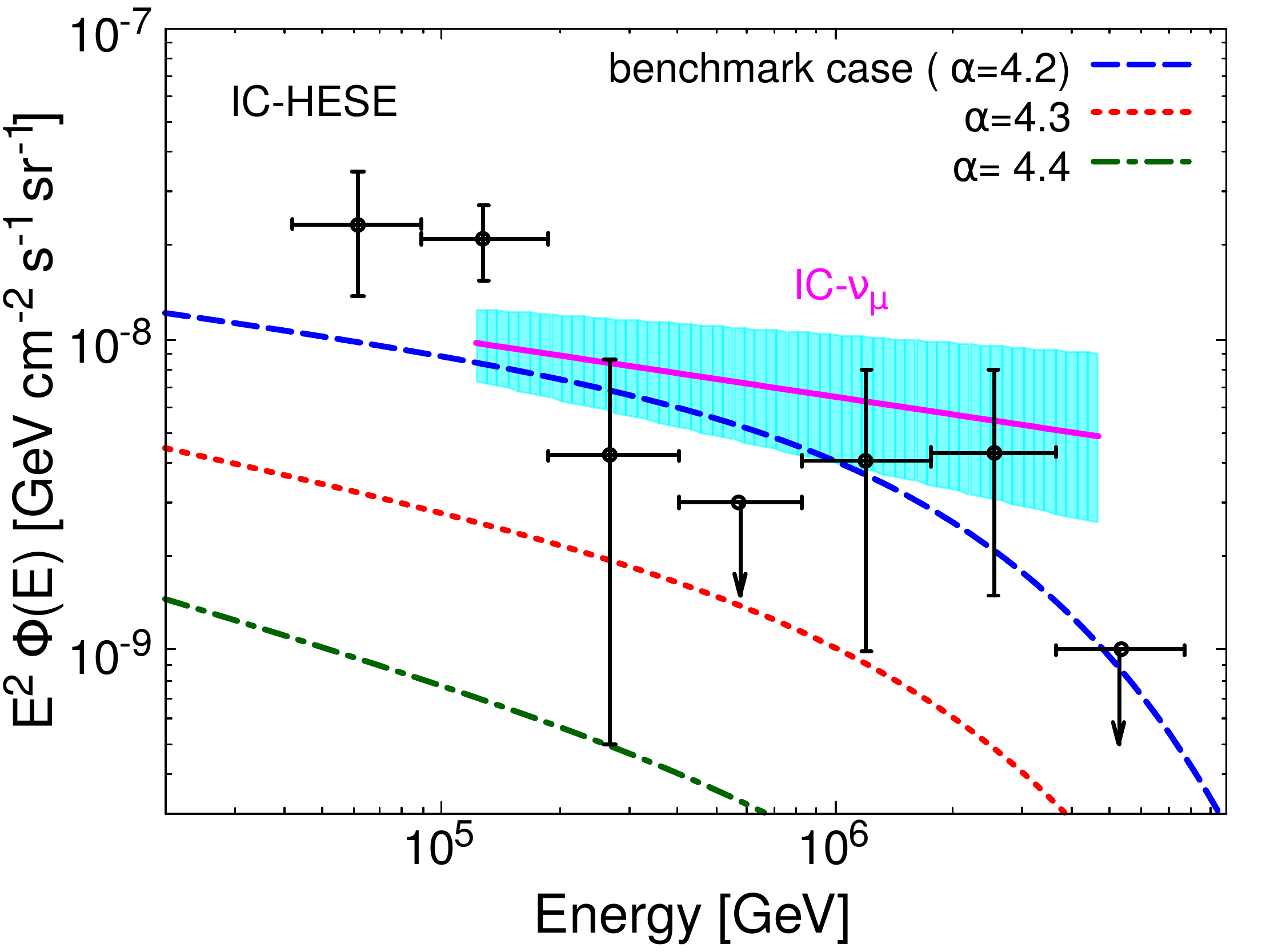} \quad
\includegraphics[width=0.44\textwidth]{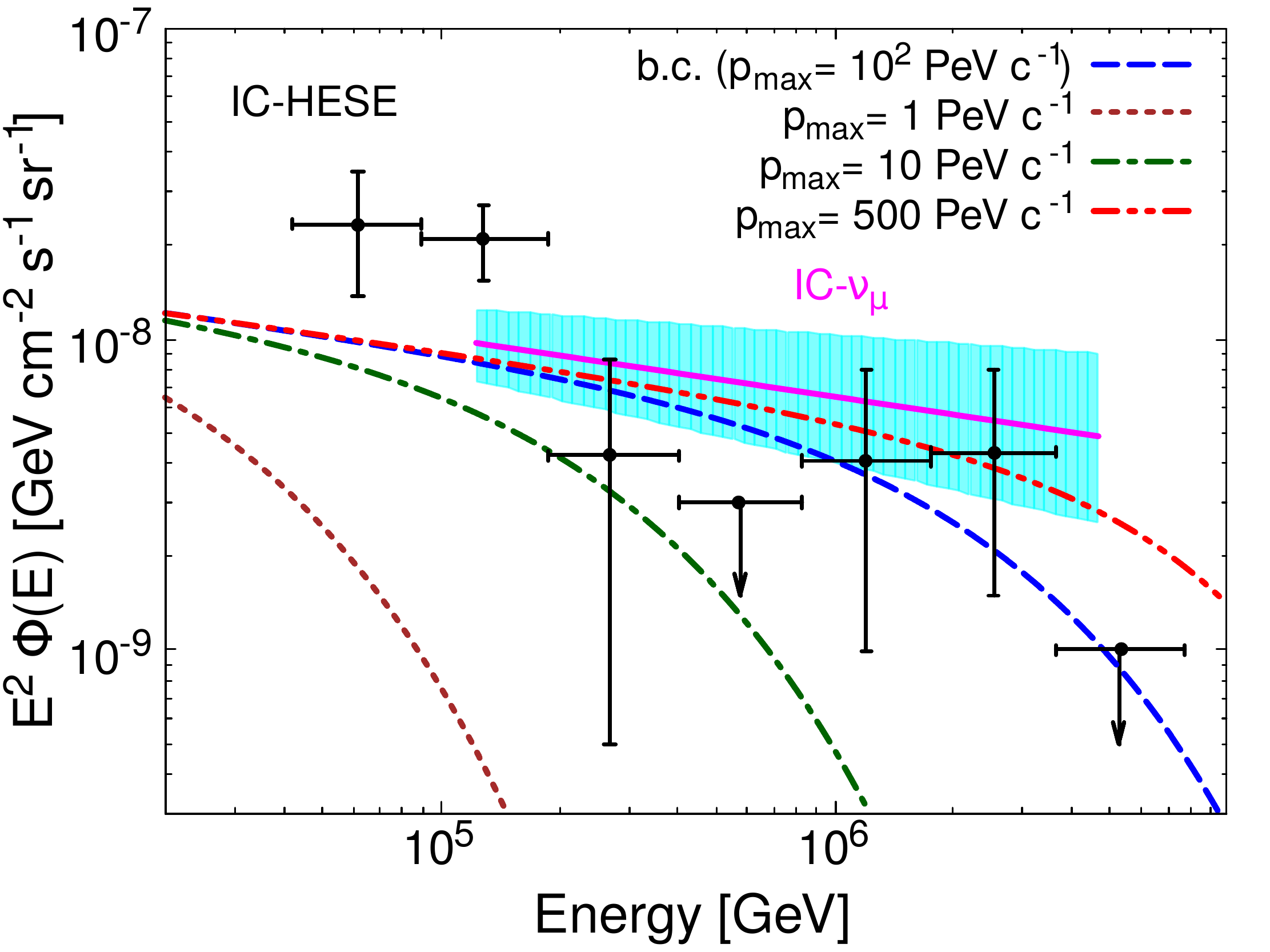} \quad
\includegraphics[width=0.44\textwidth]{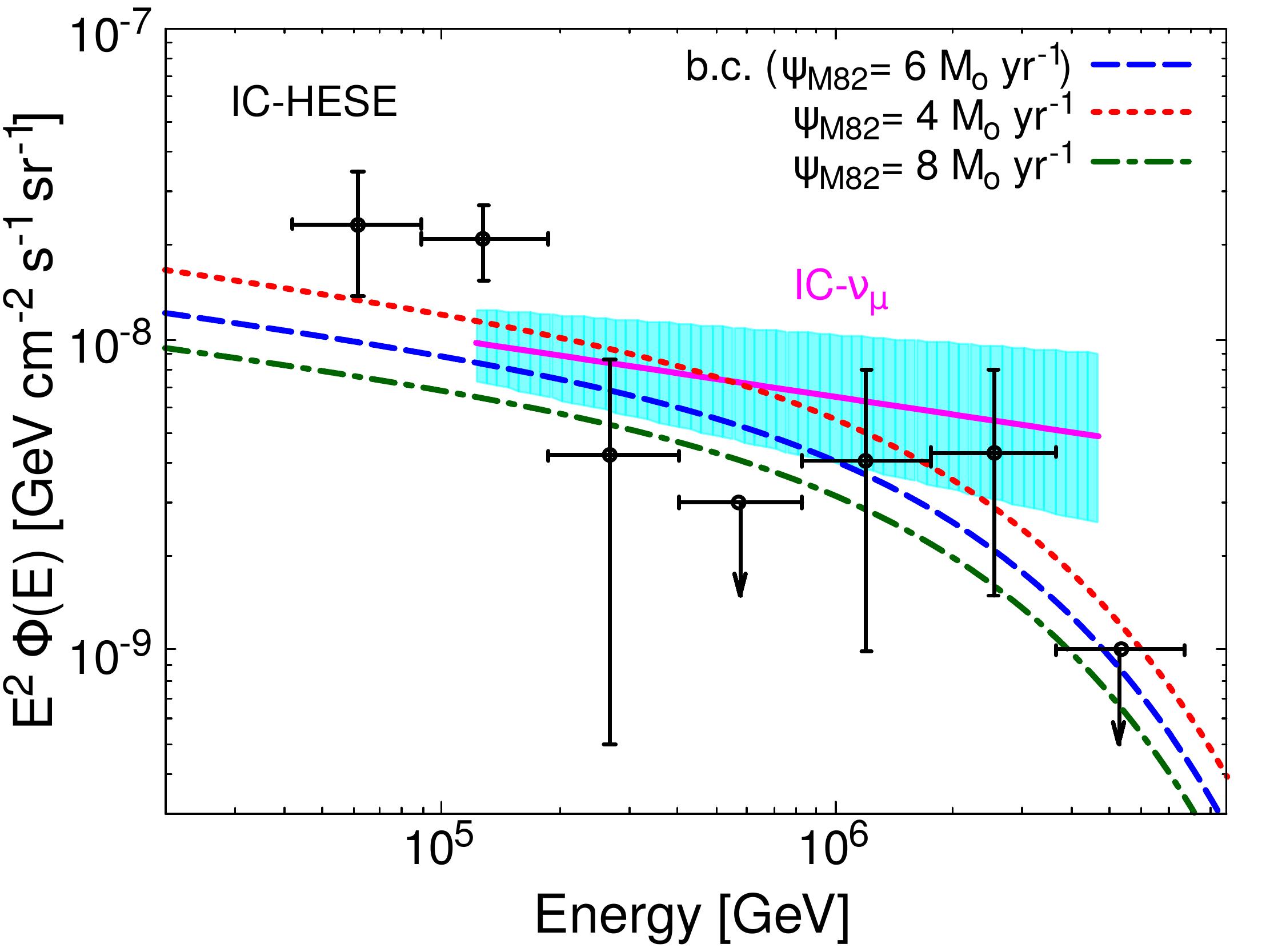}
\caption{\label{fig:Diff_param1} Single flavor neutrino flux calculated by changing the value of the three most relevant parameters, one for each panel. From top to bottom: $\alpha$, $p_{p, \rm max}$, and $\psi_{\rm M82}$. The benchmark case (b.c.) has $p_{p, \rm max}=100 ,\ \rm PeV$, $\alpha=4.2$ and $\psi_{\rm M82}= 6 \, \rm M_{\odot} \, yr^{-1}$ and it is always plotted as a dashed-blue line, while the remaining curves are as labelled. Black points are HESE neutrinos, while the shaded azure band corresponds to the through-going muon sample \citep[see][]{LASTHESE_taboada_ignacio_2018_1286919,IceCube_muons}.}
\end{figure}

In the following we comment on the dependence of the results on the choice of parameters, $p_{p, \rm max}$, $\alpha$ and $\psi_{\rm M82}$, to be used for the calculation of the diffuse gamma and neutrino flux. Any set of values for the parameters discussed below is chosen in such a way that a best fit to the multi--wavelength spectrum of M82 can be found. For the different parameters adopted, we calculated the EGB flux integrated in the $50$ GeV $- \; 2$ TeV range, reported in Tab.~\ref{tab:parameters_gamma}, and the corresponding neutrino flux, shown in the three panels of Fig. \ref{fig:Diff_param1}.

\begin{table}
\centering
\begin{tabular}{|c|c|}
\hline
 parameter set  & $\Phi_{\gamma}(E>50 \; \rm GeV)$ $\frac{10^{-10} \rm ph}{\rm  cm^2 \;s \;sr}$  \\ \hline \hline
 benchmark case & $9.2$  \\ \hline \hline
 $p_{p, \rm max}= 1 \, \rm  PeV \, c^{-1}$ & $9.3$  \\ \hline
 $p_{p, \rm max}= 10 \, \rm  PeV \, c^{-1}$ & $9.2$  \\ \hline
 $p_{p, \rm max}= 5 \cdot 10^2 \, \rm  PeV \, c^{-1}$ & $9.2$  \\ \hline
 \hline
 $\alpha=4.3$  & $6.0$  \\ \hline
 $\alpha=4.4$  & $3.5$  \\ \hline \hline
 $\psi_{\rm M82}=4 \; \rm M_{\odot}yr^{-1}$ & $12.8$  \\ \hline
 $\psi_{\rm M82}=8 \; \rm M_{\odot}yr^{-1}$ & $7.2$  \\ \hline

\end{tabular}
\caption{\label{tab:parameters_gamma} Impact on the diffuse gamma--ray flux produced by all SBNi in the energy band $50$ GeV $- \;2$ TeV changing the parameters' values with respect to the benchmark case. The benchmark case has $p_{p,\rm max}=100 \, \rm PeV$, $\alpha=4.2$ and $\psi_{\rm M82}= 6 \, \rm M_{\odot} \, yr^{-1}$. We use as upper limit for the EGB flux $9.6 \times 10^{-10} \, \rm ph \, cm^{-2} \, s^{-1} \, sr^{-1}$.}
\end{table}

The top panel of Fig.~\ref{fig:Diff_param1} illustrates the dependence of the results on the slope of the injection spectrum in the prototypical SBN, between 4.2 (benchmark case) and 4.4. Harder injection spectra appear to be ruled out by the multi--wavelength spectrum of M82 (see P18). Steeper spectra, as expected, lead to a smaller flux of neutrinos at the highest energies. The gamma--ray fluxes are also reduced correspondingly. The figure shows that the benchmark case remains the closest to the data and that steeper injection spectra cannot provide a satisfactory description of IceCube data.

The middle panel of Fig.~\ref{fig:Diff_param1} illustrates the impact of changing $p_{p, \text{ max}}$, between 1 PeV $\rm c^{-1}$ to 500 PeV $\rm c^{-1}$. The figure shows that the diffuse neutrino flux is very sensitive to the highest energy reached by accelerated protons in SBNi: for $p_{p, \rm max}=500 \, \rm PeV \, c^{-1}$ and $p_{p, \rm max}=10^2 \, \rm PeV \, c^{-1}$, the neutrino flux starts declining rather steeply at energies between $\sim 1$ PeV and a few hundred TeV, providing a satisfactory description of the IceCube data. If taken at face value, the case $p_{p, \rm max}=500 \, \rm PeV \, c^{-1}$ leads to exceeding the upper limit in the neutrino flux in the last available bin. For $p_{p, \rm max}=1 \, \rm PeV \, c^{-1}$ the decline starts at $\sim 1$ TeV, while leaving the gamma--ray flux unchanged (see Table~\ref{tab:parameters_gamma}). This latter case is clearly unable to describe IceCube data. 

Finally, the bottom panel of Fig.~\ref{fig:Diff_param1} shows the impact of a different assumption for the SFR in our M82--like prototype, within the allowed range ($4-8 \, \rm M_{\odot} \, yr^{-1}$). From to Eq.~\eqref{SB_prototype}, the dependence of the flux normalization on this parameter is linear and does not deeply affect the general result on the neutrino spectrum. On the other hand a value of $\psi_{\rm M82}\lesssim 5 \, \rm M_{\odot} \, yr^{-1}$ would lead to overproducing gamma--rays, thereby being in tension with the current EGB upper limit (see Table~\ref{tab:parameters_gamma}). 

We also checked the robustness of our results by artificially increasing $\psi^*$ up to $5 \, \rm M_{\odot} \, yr^{-1}$ and we observed that the impact on the result of our benchmark case is less than a factor $2$ in normalization. This weak dependence on $\psi^*$ can be understood analytically from Eq. \eqref{SFRF} where $\Phi_{\gamma, \nu} \propto \int d \log \psi \; \Phi_{\rm SFR}(\psi) \, f_{\gamma,\nu} (\psi)$. Accounting for the SFR-dependence of the integrand given by Eq. \eqref{SB_prototype} and Eq. \eqref{SFRF_def}, one derives an asymptotic behavior $\Phi_{\gamma,\nu} \sim \int d \psi \; \psi^{-0.6} \propto \psi^{0.4}$ (valid for values of $\psi$ much below the exponential cut-off of $\Phi_{\rm SFR}$). 

\subsection{The role of normal galaxies}

According to Eq. \eqref{min_SFR}, galaxies with a SFR lower than $\psi^*$ do not efficiently confine CRs, therefore we refer to these as normal galaxies (NGs). 
In NGs, the lack of SBNi determines a lower production rate of neutrinos and likely a diffusion--dominated cosmic--ray transport. As a consequence, the contribution of NGs to the observed neutrino flux can be expected to be negligible.
In this section, we focus on their possible contribution to the EGB. 

In order to estimate a reliable upper limit of the contribution from NGs, we assume that each of these galaxies has the same prototypical gamma--ray spectrum.
We adopt as a prototype the Milky Way Global Model (MWGB) developed by \citet{Strong-MWGM} where the HE gamma--ray flux has a spectral slope $\sim 2.7$.

%
\begin{figure}
\centering
\includegraphics[width=0.44\textwidth]{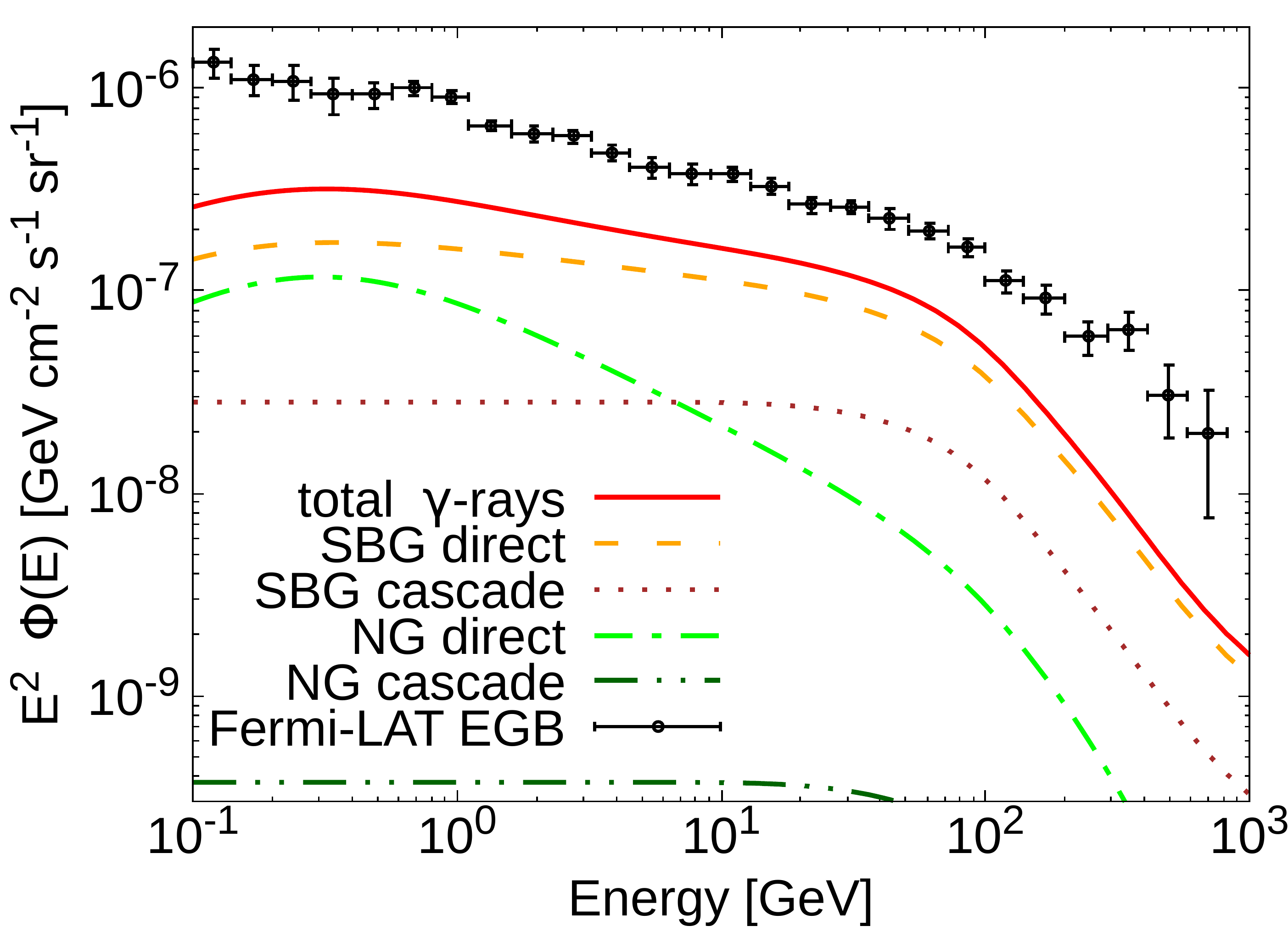} 
\caption{\label{fig:NGs} Diffuse gamma--ray background (thick red) produced by the combination of SBGs computed with the benchmark case (dashed orange) and NGs (dot-dashed green) compared with the Fermi-LAT EGB \citep[][]{Fermi-LAT<820}. The cascade contribution from SBGs (dotted brown) is shown separately from the one of NGs (dot-dot-dashed dark green). }
\end{figure}

We calculate the gamma--ray flux from NGs using  Eq.~\eqref{SFRF} where only the integration range is changed from $\psi_{\rm min}= 10^{-1.5} \, \rm M_{\odot} \, yr^{-1}$ to $\psi^*$, with $f_{\gamma}$ assumed to follow the MWGM.

In Fig.~\ref{fig:NGs} we show the results of the combination of the diffuse gamma--ray flux from SBNi, computed with the benchmark case, together with the contribution from NGs. 
We find that the contribution of NGs is only comparable to the one of SBGs below $\sim$ 1 GeV, and is reduced to $\lesssim 10 \%$ above 10 GeV.

The contribution from NGs to the photon flux at energies $\gtrsim$ 50 GeV is $\sim 6 \%$ of the one computed for SBNi (at the level of $\sim 2 \%$ of the total EGB). 

We also checked the possible impact of the contribution from NGs when their number density is computed from the IR luminosity function \citep[see][]{Gruppioni_2013_Lumin_func,Tamborra-Ando-Murase:2014} and we found compatibility within a factor $\sim 2$ with our estimate.
We conclude that the inclusion of NGs does not affect in an appreciable way the results presented in this paper.


\section{On the maximum energy of CRs in an SBN}
\label{discussion}

As discussed above, the possibility to describe the diffuse neutrino background in terms of non thermal emission from SBNi depends in a crucial way on whether the maximum energy of CR protons in sources located inside the nucleus is $\gtrsim 10$ PeV. Here we address the issue of maximum energy in somewhat more physical way: since the main feature of starburst regions is the enhanced rate of star formation and correspondingly higher rate of SN explosions, it is plausible to speculate that particle acceleration may occur at either SN shocks or at shocks developing in the stellar winds that precede the SN explosion. Since both thermal, kinetic energy and CRs are injected in a compact region (the nucleus) and typically result in the launching of fast winds, it is also plausible that particle acceleration may occur at the termination shock of such winds, as discussed by \citet{2018PhRvD..97f3010A,2018A&A...616A..57R} in connection with acceleration of ultra high energy CRs, where anisotropy observations \citep{Auger_anisotropy} and constraints based on the IGRB \citep{Liu-subankle_CR} seem to support such a positive UHECR-SBGs association.
On the other hand, such shocks are typically located away from the disc so that CRs have to propagate upwind in order to reach the dense regions where interactions are more likely to occur. Hence, in the following we concentrate on particle acceleration at SN shocks and shocks developed in individual stellar winds. 

Particle acceleration in individual stellar winds or in collective winds associated with a collection of stars in a cluster is not expected to be different in a starburst galaxy and in our Galaxy, because the energization typically takes place at shocks that are formed inside the winds. In other words, the accelerated material is the wind gas. In the Galaxy this process has been studied by~\citet{Cesarsky1983,Webb1985,bykov2011,Zirakashvili2018}. 
Based on recent gamma--ray observations~\citet{Aharonian2019} proposed that the maximum energy from the association of massive stars regions can reach the order of $\sim 1-10$~PeV. 

The situation of CR acceleration at SN shocks is more complex since acceleration is expected to take place at the forward shock, which is strongly sensitive to the conditions in the ISM in the starburst nucleus. A rule of thumb is that more turbulence on the {\it right} scales in general favors particle acceleration, where the word {\it right} here refers to the scales where particles' gyration can resonate with the scales in the turbulence. But a crucial ingredient in this type of estimates is the spectral distribution on the turbulence itself, as we discuss below. 

%

For core collapse SNe, the forward shock is expected to first propagate through the wind of the pre-supernova star, where typically the shock also reaches the Sedov phase. The beginning of the Sedov phase is also the time where the highest energy of accelerated particles is reached~\citep{schure2013}. If the situation is the same in SBNi as it is in our Galaxy, calculations of the highest energy reached at this time is somewhat below $\sim 1$ PeV, hence not sufficient for the production of neutrinos. 

If, on the other hand, the shock reaches the turbulent medium while the SN is still in the ejecta dominated phase, one could wonder if the ISM turbulence may help accelerating to higher energies. Given the large number of SN explosions, we expect the environment of a SBN to be hot and highly turbulent. As reported by \citet{Westmoquette2} for the case of M82, the ISM in an SBN is highly fragmented in clouds of different dimensions and density, mixed with star clusters, embedded in a highly pressurized ($P/k_{B} \sim 10^7$ K cm$^{-3}$) hot medium ($T \sim 10^ 6$ K and $n \sim 0.5$ cm$^{-3}$ ) \citep[see also][]{Stevens:2003ek,Yoast-Hull_M82_2013}. Moreover, given the average separation between clouds and ionizing sources, of the order of a few parsecs \citep[see again][]{Westmoquette2}, young SNRs are more likely to be found in the high temperature gas phase of low density rather than in dense clouds. 

Assuming an ejected mass of the order of $\approx 1 \rm M_{\odot}$ and a typical kinetic energy for the SN blastwave of $10^{51}$ erg one obtains an estimate for the shock speed in the ejecta dominated phase of $u_{\rm sh} \sim 10^{4}$ km s$^{-1}$. In a homogeneous medium the radius of the shock at the beginning of the Sedov phase, where the swept--up mass equals the mass of the ejecta, can be estimated to be $R_{\rm S} \approx 2.7$ pc.
\footnote{In this respect, it is interesting to notice that a radio survey of M82 done by \citet{FenechM82} detected 20 SNRs over a sample of 36 with a radius $R \lesssim 2.7 \, \rm pc$. Given the assumed SN explosion rate in M82, which ranges between 0.04 and 0.08 $\rm yr^{-1}$, this result is compatible with all those 20 SNRs being in free expansion phase.}
A common practice is to get an estimate of $E_{\rm max}$ by imposing that the the diffusion length equals a given fraction of the shock radius, usually assumed of the order of 10\% \citep[see, e.g.][]{Lagage_Cesarsky,Blasi_pmax,Ptuskin:2010zn}
\begin{equation}
\label{Emaxderiv}
    D(E_{\rm max}) \approx 0.1 \; R_{\rm S} \; u_{\rm sh} \,.
\end{equation}
Using the diffusion coefficient expressed by Eq.~\eqref{Difff1} with a Kolmogorov-like magnetic turbulence $\mathcal{F}(k)= (kL)^{-2/3}$, we obtain:
\begin{equation} 
\label{eq:E_max_K}
    E_{\max}^{(K)} \approx 7.5 \; R_{3}^3 u_{4}^3 B_{\rm mG} L_{\rm pc}^{-2} \; \rm TeV \,,      
\end{equation}
where $R_{\rm S}= 3 \; R_{3}$ pc is the Sedov radius, $u_{\rm sh}= 10^4 \; u_{4}$ km s$^{-1}$ is the shock speed, $B_{\rm mG}$ is the magnetic field in mG and $L_{\rm pc}$ is the turbulence coherence length-scale in pc. The inferred value of $E_{\max}^{(K)}$ in this case is clearly exceedingly small to have any impact on neutrino production in the IceCube energy band. Such low energies would also hard to reconcile with observation of TeV photons from M82 and NGC253.

As discussed in P18, CR transport in the SBN remains unchanged if the same magnetic field is assumed to be scale invariant, $\mathcal{F}(k)\sim \rm constant$, which reflects in Bohm diffusion in terms of particle diffusion. This assumption has however tremendous implications in terms of maximum energy of accelerated particles. In the Bohm limit the condition \eqref{Emaxderiv} reads:
\begin{equation}
    E_{\rm max}^{(B)} \approx 30 \; R_{3} \, u_{4} \, B_{\rm mG} \; \rm PeV.
\end{equation}
The average magnetic field for starburst nuclei regions is typically inferred $\gtrsim 100$ $\mu$G and might reach a few mG in the nuclear regions of ULIRG like Arp 220 \citep[][]{Torres_2004,Yoast-Hull-ARP}. In this case a maximum energy in the range $50-100$ PeV, optimal for neutrino production in the PeV region, can be seen as a manifestation of the high level of turbulence in the nucleus, where SN explosions occur. 

One might also wonder if CR induced instabilities may locally enhance the level of turbulence, thereby reducing the acceleration time. These mechanisms are crucial in SNe exploding in our Galaxy, if they are assumed to be the sources of CRs up to the knee. Two such instabilities are especially important:  the non--resonant Bell instability (NRI) \citep{Bell2004,Amato_Blasi_CR} and the amplification driven by CR pressure gradients in an inhomogeneous plasma~\citep{Drury_Turbulence}.

As discussed by \citet{Bell2004}, the NRI is induced by the CR current when the condition $P_{\rm CR} u_{\rm sh}/c > B_0^2/4\pi$ is fulfilled. Since the CR pressure in the shock region is a fraction $\xi_{\rm CR}= 0.1  \; \xi_{0.1} < 1$ of the upstream fluid ram pressure $\rho_0 u_{\rm sh}^2$, the last condition translates to an upper limit for the initial magnetic field which reads: 
\begin{equation}
    	B_0  \ll 260  \, \sqrt{ \xi_{\rm 0.1} n_0} \, u_4^{3/2} \,  {\rm \mu G} \,.
\label{eq:Bell_sat}	
\end{equation}
For typical conditions in a SBN, this condition is either barely satisfied or not satisfied at all. Even the instability were excited, the saturated magnetic field is, at most, of the same order as the RHS of Eq. \eqref{eq:Bell_sat}, thereby making the role of this instability in SBNi very limited. 

If density inhomogeneities are present upstream, then the magnetic field can also go through turbulent amplification \citep[see e.g.][]{Drury_Turbulence,2009ApJ...707.1541B}, provided the following condition is fulfilled:
\begin{eqnarray}
    u_{\rm sh} > v_{\rm A} \sqrt{4 \pi} \frac{1}{\xi_{\rm CR}} 
    		\frac{1}{(\delta \rho/\rho_0)} \hspace{3cm} \nonumber \\
    \Rightarrow \;  B <  0.13 \,\, \xi_{0.1} 
    		\sqrt{n_0} \, \frac{\delta\rho}{\rho_0}\, u_4 \,  {\rm mG} \,,
    \label{Drury}
\end{eqnarray}
and $\delta\rho/\rho_0$ denotes the strength of the density fluctuations on scales smaller than the size of the CR precursor upstream of the shock. With typical values of the parameters and assuming $\delta\rho/\rho_0\sim 1$ an amplification of the magnetic field by about an order of magnitude appears plausible, so as to drive an original magnetic field of $\sim 100 \, \mu \rm G$ up to the mG level. Notice that the pressure gradient that causes the instability to grow assumes that particle acceleration is already efficient, so as to produce a pronounced CR precursor. Hence this mechanism should be considered mainly as responsible for an increase in the maximum energy of the accelerated particles, while the whole acceleration is probably bootstrapped either due to the ambient magnetic field of the SBN or due to the magnetic field produced through NRI. 

We conclude that particle acceleration up to $50-100$ PeV in sources located in SBNi is plausible although by no means trivial. Observational guidance in assessing this problem is, at this point, crucial.

\section{Conclusions}
\label{Section_conclusions}

We have studied the contribution of SBNi to the diffuse flux of gamma rays and HE neutrinos assuming a SFRF approach based on observations \citep[][]{Gruppioni_2015_2}. Requirements on the confinement of protons helped us in defining the properties needed for a galactic nuclear region to be an efficient CR calorimeter. The fluxes of gamma rays and neutrinos then become functions of only injection parameters ($\alpha$ and $E_{\rm max}$) and lose their dependence on other environmental quantities. Relying on the confinement requirement, we built an SBN prototype model based on CR transport and fit the parameters to the multifrequency electromagnetic spectrum of M82. The model also accounts for gamma--ray absorption both inside the source and during transport on cosmological scales. 

If one takes the M82--like prototypical model at face value, the diffuse neutrino flux contributed by SBNi is in agreement with the observed IceCube neutrino flux for $E\gtrsim 200$ TeV, while at the same time accounting for about $\sim 40\%$ of the EGB observed by Fermi--LAT at energies $>50$ GeV. This contribution, together with that of blazars would therefore saturate the observed EGB. 

The multifrequency spectrum of M82 strongly constrains the spectrum of CRs injected in its SBN. This is due to the fact that CRs loose their energy inside the SBN, so that the equilibrium spectrum approximately reproduces the source spectrum (the only exception is due to the weak energy dependence of the inelastic cross section). This implies that once the spectral shape of the neutrino spectrum is fixed at the highest energies by fitting observations, the low energy part is also fixed. As a consequence, at energies below $\sim 200$ TeV the predicted neutrino flux is about a factor $\sim 2$ below the data. This conclusion might be affected by the assumption that SBNi can be modelled as M82-like objects with suitable rescaling of their CR content, but at present this appears to be a well justified assumption. 

It is also possible that the observed neutrino flux in the lowest energy bins may contain a contribution from other sources, from our own Galaxy \citep[see e.g.][]{Taylor_Neutrinos,NERONOV201660,Gal-neutrinos,Blasi_Amato19} or might even be affected by a contamination due to atmospheric neutrinos \citep{Mascaretti_Blasi,Mascaretti_Vissani_prompt}.

In this sense, our results support the picture based on a multi--component interpretation of the global diffuse flux measured by IceCube, as suggested by \cite{Palladino_multi1,Palladino-multicomponent:2018}. In fact our computed flux of neutrinos of starburst origin is in good agreement with the results of \citet{Palladino-starburst:2018}. The main difference among these conclusions and those of \citet{Bechtol-Ahlers:2015} is, we believe, based upon the weight to associate to the first two data points of IceCube. 

The robustness of our results has been tested by exploring the parameter space in terms of slope of the injected CR spectrum and maximum energy, SFR of the prototypical SBN and the minimum SFR above which a SBN can be considered as an efficient neutrino factory.

While the level of SFR, if chosen within the astrophysical uncertainties, only marginally affects our general conclusions, changing the CR spectrum can modify our final statements on neutrinos. The maximum energy of accelerated protons is required to be $\gtrsim 50 \, \rm PeV$. Smaller energies fails to explain the IceCube data, while leaving our conclusions on the diffuse gamma rays unchanged. The slope of the injection spectrum is fixed to be $4.2$ because of the constraint to fit the multifrequency spectrum of the prototypical M82-like galaxy. Steeper spectra fail to reproduce neutrino observations. 
%
%
As discussed in P18, if the level of turbulence in the SBN is low or moderate, the diffusion coefficient can be appreciably larger than assumed here. 

This would imply the existence of a transition from loss--dominated (calorimetry) to diffusion dominated transport, which would reflect into a steepening of the spectrum of secondaries.
In the context of this model it would be difficult to describe the photon spectrum of our stereotypical SBN.


We briefly discussed the issue of accelerating particles up to $\sim 50-100$ PeV necessary to explain neutrino observations. We found that, though not unrealistic, it is not trivial to reach such high energies in sources inside SBNi. The high level of turbulence expected in the SBN as due to repeated SN explosions, possibly in combination with some type of turbulent amplification of the magnetic field upstream might in fact lead to push $E_{max}$ up into the interesting range, $\gtrsim 50$ PeV.


\section*{Acknowledgements}

EP is very grateful to Pierluigi Monaco for his precious help and suggestions. We thanks G. Pagliaroli, A. Palladino and F. Vissani for stimulating discussions at various stages of the development of the manuscript. PB and GM acknowledge the support received through Grants ASI/INAF n. 2017-14- H.O and SKA-CTA-INAF 2016.





\bibliographystyle{mnras}
\bibliography{references}




\appendix

\section{Redshift effect on flux}

\label{app:calculation}
The arrival flux of gamma rays and neutrinos observed at energy $E_{0}$ can be easily calculated from the number of photons or neutrinos produced at the source with energy $E= E_0 (1+z) $ as:
\begin{eqnarray}
\label{Az}
    \frac{df_{\gamma}(E_0)}{dE_0} = \frac{dN_{\gamma}(E_0)}{dS \; dt_0 \; dE_0}= (1+z)^2 \frac{\varphi_{\gamma}(E_0 [1+z])}{4 \pi d_{\rm L}^2(z)} e^{-\tau_{\gamma \gamma}(E_0,z)} \\
    \frac{df_{\nu}(E_0)}{dE_0} = \frac{dN_{\nu}(E_0)}{dS \; dt_0 \; dE_0}= (1+z)^2 \frac{\varphi_{\nu}(E_0 [1+z])}{4 \pi d_{\rm L}^2(z)}
    \label{Bz}
\end{eqnarray}
where $z$ is the redshift, $d_L(z)$ the luminosity distance and $\tau_{\gamma \gamma}$ is the gamma--ray optical depth as computed by \cite{Franceschini_EBL_erratum}.

\section{Derivation of the asymptotic behavior of electromagnetic cascade spectrum}
\label{Appendix C}

A high--energy gamma--ray escaping a source has to traverse a further opaque medium before reaching the Earth. The intergalactic medium (IGM) filled by photons of the cosmic microwave background (CMB) and by extragalactic background light (EBL) photons, can be highly opaque already at $\sim 10^2$ GeV. If a gamma--ray is absorbed its energy is reprocessed in the electromagnetic cascade. This process, as discussed in detail in \citet{Berezinsky_em_cascade:2016}, consists in a transfer of energy from very energetic gamma rays and leptons into photons, whose spectrum is displaying universal features.  Gamma rays transfer their energy to leptons via pair production and leptons upscatter low energy target radiation producing high--energy photons. This process can be divided in three main phases: leading particle regime, cascade multiplication and low-energy regime, and it ends when the interaction length for electrons is longer than their distance to the Earth. 
The approach we adopt for the computation of the cascade spectrum is inspired by the analytic approach proposed in \citet{Berezinsky_em_cascade:2016}, where it is assumed that the cascade has enough time to fully develop and that the low--energy photon background can be approximated by delta functions: 
\begin{equation}
    n_{ph}(z,\epsilon_1)= \mathcal{C}_{\rm CMB}(z) \delta\big[\epsilon_1-\epsilon_{\rm CMB}(z)\big] + \mathcal{C}_{\rm EBL}(z) \delta\big[\epsilon_1-\epsilon_{\rm EBL}(z)\big],
\end{equation}
where $\mathcal{C}_j$ and $\epsilon_j$ are the redshift-dependent normalization constant and the energy of the peak of the photon field $j=$ ${\rm EBL}$ or ${\rm CMB}$.

The leading particle regime takes place at the highest energies where, as discussed in \citet{Aharonian_Book}, the pair production transfers the energy of the parent photons preferentially to only one of the two leptons, and the inverse Compton scattering takes place in Klein-Nishina regime. During this phase both pair production and inverse Compton involve the CMB because of its dominant number density. When the energy of a gamma-ray is lower than $\mathcal{E}^{\rm CMB}_{1}(z)= m_e^2 c^4/ \epsilon_{\rm CMB}(z)$, the cascade enters its second stage, namely the cascade multiplication, where the EBL becomes the target photon field for the pair-production, whereas the inverse Compton keeps on upsattering the most numerous CMB, in Thomson regime from now on. In the cascade multiplication, the upscattered radiation is reprocessed in the cascade if it is energetic enough to interact again with the EBL, namely if $E_1 \geq \mathcal{E}_{\gamma,1}(z) = m_e^2 c^4/\epsilon_{\rm EBL}(z)$, otherwise the Universe becomes transparent and it reaches us as a part of the high energy branch of the cascade spectrum. The electron-positron pairs produced during the cascade multiplication supply a new generation of gamma-rays that can keep on the cascade down to the critical Lorentz factor $\gamma_{min}=\mathcal{E}_{\gamma,1}(z)/2m_ec^2$, where the upscattered photons have the critical energy $\mathcal{E}_{\rm X,1}(z)= 4 \gamma^2_{min} \epsilon_{\rm CMB}(z)/3= m_e^2c^4 \epsilon_{\rm CMB}(z)/3 \epsilon_{\rm EBL}^2(z)$.

The asymptotic shape of the universal spectrum of the electromagnetic cascade can be determined considering that the overall number of electrons of energy $E_e$ appearing in the entire cascade history $n_e(E_e)=\int dt \; q_e^{\rm cas}(E_e^{*},t) \delta[E_e^{*}-E_e] $ is constant for $E_e \leq E_{e,min} = \gamma_{\rm min} m_ec^2$, and proportional to $E_e^{-1}$ at higher energy. The inverse Compton origin of the cascade photons imply $E_{\gamma} \propto E_e^2$, therefore the energy conservation can be applied as follows:
\begin{equation}
    \label{E_cons_cascade}
    E_{\gamma} d n_{\gamma}(E_{\gamma}) \propto n_e(E_e) dE_e,
\end{equation}
where $dE_e$ is the energy lost by electrons of energy $E_e$, and $d n_{\gamma}$ is the number of photons upscattered at the energy $E_{\gamma}$. Considering the relation between $E_e$ and $E_{\gamma}$, and the energy dependence of $n_e$, Eq. \ref{E_cons_cascade} reads
\begin{align}
    \frac{d n_{\gamma}(E_{\gamma})}{dE_{\gamma}} \propto \frac{dE_e}{dE_{\gamma}} \frac{1}{E_{\gamma}} n_e(E_e) \propto \frac{1}{E_{e}^{3}} 
    \begin{cases} 
    \text{const} & E_e \leq E_{e,\text{min}} \\
    E_e^{-1} &  E_e \geq E_{e,\text{min}} 
    \end{cases}
    \\
    \propto
    \begin{cases}
    E_{\gamma}^{-3/2} & E_{\gamma} \leq \mathcal{E}_{\rm X} \\
    E_{\gamma}^{-2} & E_{\gamma} \leq \mathcal{E}_{\rm X} \leq \mathcal{E}_{\gamma} \\
    0 & E_{\gamma} \geq \mathcal{E}_{\gamma}
    \end{cases}
\end{align}
where the critical energies $\mathcal{E}_{\rm X}$ and $\mathcal{E}_{\gamma}$, and so the spectrum $d n_{\gamma}/dE_{\gamma}$ are redshift dependent.

In our calculation, instead of assuming a sharp cut off at $\mathcal{E}_{\gamma}$, we consider the proper suppression due to the EBL+CMB optical depth $\tau_{\gamma \gamma}(E,z)$. 
The cascade normalization is computed by assuming that it fully develops at the same redshift where cascading gamma-rays are emitted and fully converting the energy of absorbed photons in energy density of the cascade.
At fixed redshift the cascade energy content is computed as:
\begin{equation}
    \int dE \; E \; f_{\rm cas}(E,z) = \mathcal{N}_{\rm SBG}(z) \int dE \; E \; f_{\rm SBG}(E,z) [1-e^{-\tau_{\gamma \gamma}(E,z)}] \, ,
    \label{casc norm}
\end{equation} 
where $\mathcal{N}_{\rm SBG}(z)$ is the number of SBGs at a given redshift $z$, $f_{\rm cas}$ is the cascade photon flux and $f_{\rm SBG}$ is the emitted flux from the starburst.

\bsp	
\label{lastpage}
\end{document}